\documentstyle[aps,12pt,epsf,eufrak]{revtex}

\newcommand{\Rsub}{\rm\scriptscriptstyle}

\def\slashchar#1{\setbox0=\hbox{$#1$}           
   \dimen0=\wd0                                 
   \setbox1=\hbox{/} \dimen1=\wd1               
   \ifdim\dimen0>\dimen1                        
      \rlap{\hbox to \dimen0{\hfil/\hfil}}      
      #1                                        
   \else                                        
      \rlap{\hbox to \dimen1{\hfil$#1$\hfil}}   
      /                                         
   \fi}                                         %
\thicklines
\begin{document}
\begin{center}
{\large\bf  Heavy quark potential in the static limit of QCD}\\
\vspace*{4mm}
{\sf V.V.Kiselev}\footnote{E-mail: kiselev@th1.ihep.su}, 
{\sf A.E.Kovalsky}\footnote{Moscow Institute of Physics and Technology,
Dolgoprudny, Moscow region.\\ \hspace*{4mm}E-mail:
kovalsky\underline{~}a\underline{~}e@mail.ru},\\
\vspace*{3mm}
Russian State Research Center "Institute for High Energy Physics",\\
Protvino, Moscow Region, 142284 Russia\\
\vspace*{3mm}
and\\
{\sf A.I.Onishchenko}\footnote{E-mail: onischen@heron.itep.ru}\\
\vspace*{3mm}
{Institute for Theoretical and Experimental Physics,\\ 
B. Cheremushkinskaja, 25, Moscow, 117259 Russia}
\end{center}
\vspace*{2mm}
\begin{abstract}
Following the procedure and motivations developed by Richardson, Buchm\"uller
and Tye, we derive the potential of static quarks consistent with both the
three-loop running of QCD coupling constant under the two-loop perturbative
matching of  V and $\overline{\rm MS}$ schemes and the confinement regime at
long distances. Implications for the heavy quark masses as well as the
quarkonium spectra and leptonic widths are discussed.
\end{abstract}

\vspace*{1cm}

\vspace*{1cm} 
\section{Introduction}
The potential of static heavy quarks illuminates the most important features of
QCD dynamics: the asymptotic freedom and confinement. Trying to study subtle
electroweak phenomena in the heavy quark sector of Standard Model, we need
quite an accurate quantitative understanding of effects caused by the strong
interactions. In addition to the perturbative calculations for hard
contributions, at present there are three general approaches to get a
systematic description of how the heavy quarks are bound into the hadrons and
what are the relations between the measured properties of such the hadrons and
the characteristics of heavy quarks as relevant to the electroweak interactions
and QCD. These approaches are the Operator Product Expansion (OPE) in the
inverse powers of heavy quark mass, the Sum Rules (SR) of QCD and the Potential
Models (PM) for the systems containing the heavy quarks by exploring various
approximations of Bethe-Salpeter equation with the static potential treated in
the framework of effective theory with a power counting in terms of powers of
the inverse heavy quark mass. The first method is usually exploited in the
inclusive estimates, while the second and third techniques are the frameworks
of exclusive calculations. The important challenge is a consistency of
evaluations obtained in such the ways, that requires the comparative analysis
of calculations. A wide variety of systems and processes for the analysis
provides a more complete qualitative and quantitative understanding of heavy
quark dynamics.

In the leading order of perturbative QCD at short distances and with a linear
confining term in the infrared region, the potential of static heavy quarks was
considered in the Cornell model \cite{Corn}, incorporating the simple
superposition of both asymptotic limits (the effective coulomb and string-like
interactions). The observed heavy quarkonia posed in the intermediate
distances, where both terms are important for the determination of mass
spectra. So, the phenomenological approximations of potential (logarithmic one
\cite{log} and power law \cite{Mart}), taking into account the regularities of
such the spectra, were quite successful \cite{pmQR}, while the quantities more
sensitive to the global properties of potential are the wave functions at the
origin as related to the leptonic constants and production rates. So, the
potentials consistent with the asymptotic freedom to one and two loops as well
as the linear confinement were proposed by Richardson \cite{Richard},
Buchm\"uller and Tye \cite{BT}, respectively. Technically, using a given scheme
of regularization, say, $\overline{\rm MS}$, one has to calculate the
perturbative expansion for the potential of static quarks. This potential can
be written down as the coulomb one with the running coupling constant in the
so-called  V scheme. Thus, the perturbative calculations provide us with the
matching of $\overline{\rm MS}$ scheme with  V-one. The $n$ loop running of
$\alpha_s^{\overline{\Rsub MS}}$ requires the $n-1$ loop matching to
$\alpha_{\Rsub V}$. Note, that initial two coefficients of corresponding
$\beta$ functions are scheme and gauge independent, while others generally
depend. With the dynamical fields integrated out, the V scheme is defined in
terms of the action depending on the static sources (the distance $r$), so that
its $\beta$ function is gauge invariant. The motivation by Buchm\"uller and Tye
was to write down the $\beta$ function of $\alpha_{\Rsub V}$ consistent with
two known asymptotic regimes at short and long distances. They proposed the
function, which results in the effective charge determined by two parameters,
only: the perturbative parameter is the scale in the running of coupling
constant at large virtualities and the nonperturbative parameter is the string
tension. The necessary inputs are the coefficients of $\beta$ function. Two
loop results and the one loop matching condition were available to the moment.
Recently, the progress in calculations has provided us with the two loop
matching of  V and $\overline{\rm MS}$ schemes \cite{Peter,Schroed}, that can
be combined with the three loop running of $\alpha_s^{\overline{\Rsub MS}}$.
So, the modification of Buchm\"uller--Tye (BT) potential of static quarks as
dictated by the current status of perturbative calculations is of great
interest. Moreover, to the moment two peculiar questions become open. First,
the asymptotic perturbative expansion of BT $\beta$ function to the third order
results in the three loop coefficient, which is wrong even in its sign. Second,
the elaborated $\Lambda_{\overline{\Rsub MS}}$ parameter by BT is in a deep
contradiction with the measured value \cite{PDG}. To clarify the situation, we
are tending to derive the static quark potential consistent with the state of
the art.

Thus, our motivation is to combine high order multi-loop calculations of the
perturbative static potential \cite{Peter,Schroed} with the string tension
ansatz. So, we improve the perturbative input for the potential model in order
to remove the contradiction between the modern high energy data on the QCD
coupling constant and the description of heavy quark potential in the framework
of one-loop Buchm\"uller--Tye model, which accepts extremely high value of
coupling constant evolved to the $Z$ mass scale. In other words, if we accept
the current normalization of coupling constant and introduce its value into the
Buchm\"uller--Tye approach to the one-loop potential, then we get the
contradiction of such the potential with the heavy quarkonium mass spectra,
certainly, since we find about 200 MeV smaller splitting between the $1S$ and
$2S$ levels in comparison with experimental 580 MeV. This discrepancy cannot be
removed by the modification of nonperturbative part in the potential with no
contradiction with the data on the slope of Regge trajectories. Therefore, the
modification of perturbative input for the model of static potential in QCD is
meaningful in this sense even when the nonperturbative contribution is
conserved in the old string tension form. So, the significant improvement of
perturbative $\beta$ function for the charge in the coulomb potential is
combined with the consequent evolution from high virtualities to low ones with
taking into account the influence of nonperturbative term on the evolution,
that becomes essential numerically below the scale of 4 GeV. 

We have to emphasize that to the moment of paper by Buchm\"uller and Tye
\cite{BT} a theory for the heavy nonrelativistic $Q\bar Q$ pair did not exist.
So, the phenomenological derivation of the static potential including
perturbative short-distance and non-perturbative long-distance elements made by
BT was all one could do. At present, at least for very heavy quarks, such a
theory does exist in the form of pNRQCD \cite{pNRQCD} and vNRQCD \cite{vNRQCD},
and we address the comparison of static potential model developed in this work
with these sound theoretical approaches in QCD to the physics of heavy
quarkonium.

Another aspect of this work is devoted to the heavy quark masses. After the
potential is given, the heavy quark masses incorporated in the corresponding
Schr\"odinger equation determine the heavy quarkonium spectra with
no ambiguity\footnote{We deal with the so-called spin-averaged spectra, since
the consideration of spin-dependent splitting involves some additional
parameters beyond the static potential.}. These masses involved in the
potential model are denoted by $m_Q^{\Rsub V}$. Such the mass should be
distinguished from the pole mass which is a purely perturbative concept defined
unambiguously at each order of perturbation theory through the pole of the
perturbative heavy quark propagator. Thus, we need to test the consistency
of estimates for the masses in the QCD potential of static quarks and in SR,
say.

In Section 2 we generalize the BT approach to three loops and derive the
static potential of heavy quarks. The numerical values of potential parameters
and their consistency with the relevant quantities are considered. The
implications for the heavy quark masses, spectra of heavy quarkonia
and leptonic constants are discussed in Section 3. The obtained results are
summarized in Conclusion.

\section{QCD and potential of static quarks}
In this section, first, we discuss two regimes for the QCD forces between the
static heavy quarks: the asymptotic freedom and confinement. Second, we
formulate how they can be combined in a unified $\beta$ function obeyed both
limits of small and large QCD couplings.

\subsection{Perturbative results at short distances}
The static potential is defined in a manifestly gauge invariant way by means of
the vacuum expectation value of a Wilson loop \cite{Su},
\begin{eqnarray}
\label{def_WL}
V(r) &=& - \lim_{T\rightarrow\infty} 
\frac{1}{iT}\, \ln \langle{\cal W}_\Gamma\rangle \;, \nonumber\\
{\cal W}_\Gamma &=& \widetilde{\rm tr}\, 
{\cal P} \exp\left(ig \oint_\Gamma dx_\mu A^\mu\right) \;.
\end{eqnarray}
Here, $\Gamma$ is taken as a rectangular loop with time extension 
$T$ and spatial extension $r$. The gauge fields $A_\mu$ are 
path-ordered along the loop, while the color trace is normalized 
according to $\widetilde{\rm tr}(..)={\rm tr}(..)/{\rm tr}1\!\!1\,$.

Generally, one introduces the  V scheme of QCD coupling constant by the
definition of QCD potential of static quarks in momentum space as follows:
\begin{equation}
 V({\bf q}^2)  =  -C_F\frac{4\pi\alpha_{\Rsub V}({\bf q}^2)}{{\bf q}^2},
\end{equation}
while $\alpha_{\Rsub V}$ can be matched with $\alpha_{\overline{\Rsub MS}}$
\begin{eqnarray}
\alpha_{\Rsub V}({\bf q}^2) & = & \alpha_{\overline{\Rsub MS}}(\mu^2)
\sum_{n=0}^\infty \tilde{a}_n(\mu^2/{\bf q^2})
\left(\frac{\alpha_{\overline{\Rsub MS}}(\mu^2)}{4\pi}\right)^n
\label{orig} \\ &=& 
\alpha_{\overline{\Rsub MS}}({\bf q}^2)
\sum_{n=0}^\infty a_n\left(\frac{\alpha_{\overline{\Rsub MS}}({\bf q}^2)}
{4\pi}\right)^n. \label{vms}
\end{eqnarray}
At present, our knowledge of this expansion\footnote{On a possible peculiar
behaviour in the expansion see ref.\cite{Su}.} is restricted by
\begin{equation}
a_0=\tilde a_0=1,\quad\quad
a_1 = \frac{31}{9}C_A - \frac{20}{9}T_Fn_f,\quad\quad
\tilde a_1 = a_1 +\beta_0\ln\frac{\mu^2}{\bf q^2},
\end{equation}
which is the well-known one-loop result, and the recent two-loop calculations
\cite{Peter,Schroed}, which gave
\begin{eqnarray}
a_2 &=& \left(\frac{4343}{162}+4\pi^2-\frac{\pi^4}{4}+\frac{22}{3}\zeta(3)
\right)C_A^2 -\left(\frac{1798}{81}+\frac{56}{3}\zeta(3)\right)C_AT_Fn_f
\nonumber \\
&&
-\left(\frac{55}{3}-16\zeta(3)\right)C_FT_Fn_f +
\left(\frac{20}{9}T_Fn_f\right)^2,  \\
\tilde a_2 &=& a_2 + \beta_0^2\ln^2\frac{\mu^2}{\bf q^2}
+(\beta_1+2\beta_0 a_1)\ln\frac{\mu^2}{\bf q^2}.
\end{eqnarray}
We have used here the ordinary notations for the SU($N_c$) gauge group: $C_A =
N_c$, $C_F = \frac{N_c^2-1}{2 N_c}$, $T_F = \frac{1}{2}$. The number of active
flavors is denoted by $n_f$.

After the introduction of ${\EuFrak a} =\frac{\alpha}{4 \pi}$, the $\beta$
function is actually defined by 
\begin{equation}
\frac{d {\EuFrak a}(\mu^2)}{d\ln\mu^2} = \beta({\EuFrak a})
  = - \sum_{n=0}^\infty \beta_n \cdot {\EuFrak a}^{n+2}(\mu^2),
\end{equation}
so that $\beta_{0,1}^{\Rsub V}=\beta_{0,1}^{\overline{\Rsub MS}}$ and
\begin{eqnarray}
\beta_2^{\Rsub V} &=&
\beta_2^{\overline{\Rsub MS}}-a_1\beta_1^{\overline{\Rsub MS}}
+ (a_2-a_1^2)\beta_0^{\overline{\Rsub MS}} \\
& = & \Big(\frac{618+242\zeta(3)}{9}+\frac{11(16\pi^2-\pi^4)
    }{12}\Big)C_A^3 \nonumber \\ &&
    -\Big(\frac{445+704\zeta(3)}{9}+\frac{16\pi^2-\pi^4}{3}\Big)C_A^2T_F
    n_f \nonumber \\ &&
    + \frac{2+224\zeta(3)}{9}C_A(T_Fn_f)^2
    -\frac{686-528\zeta(3)}{9}C_AC_FT_Fn_f
    \nonumber \\ &&
    +2C_F^2T_Fn_f+\frac{184-192\zeta(3)}{9}C_F(T_Fn_f)^2.
\label{b2v}
\end{eqnarray}
The coefficients of $\beta$ function, calculated in the $\overline{\rm MS}$
scheme \cite{TVZ80} are given by
\begin{eqnarray}
\beta_0^{\overline{\Rsub MS}} & = & \frac{11}{3} C_A -\frac{4}{3} T_F n_f,\\
\beta_1^{\overline{\Rsub MS}} & = & \frac{34}{3} C_A^2 - 4 C_F T_F n_f
-\frac{20}{3} C_A T_F n_f,\\
\beta_2^{\overline{\Rsub MS}} & = & \frac{2857}{54}C_A^3+2C_F^2T_Fn_f
    -\frac{205}{9}C_AC_FT_Fn_f-\frac{1415}{27}C_A^2T_Fn_f \nonumber \\
   && +\frac{44}{9}C_F(T_Fn_f)^2+\frac{158}{27}C_A(T_Fn_f)^2.
\end{eqnarray}
The Fourier transform results in the position-space potential \cite{Peter}
\begin{eqnarray}
 V(r) & = & -C_F\frac{\alpha_{\overline{\Rsub MS}}(\mu^2)}{r}\Bigg(
      1 + \frac{\alpha_{\overline{\Rsub MS}}(\mu^2)}{4\pi}\Big(
          2\beta_0\ln(\mu r^\prime)+a_1\Big) \nonumber \\
   && + \Big(\frac{\alpha_{\overline{\Rsub MS}}(\mu^2)}{4\pi}\Big)^2 \Big(
      \beta_0^2(4\ln^2(\mu r^\prime)+\frac{\pi^2}{3}) \label{Vr} \\
   && \quad +2(\beta_1+2\beta_0a_1)\ln(\mu r^\prime)+a_2\Big) + \ldots \Bigg)
      \nonumber
\end{eqnarray}
with $r^\prime\equiv r\exp(\gamma_E)$. Defining the new running coupling
constant, depending on the distance,
\begin{equation}
  V(r) = -C_F\frac{\bar\alpha_{\Rsub V}(1/r^2)}{r}.
\end{equation}
we can calculate its $\beta$-function from (\ref{Vr}), so that \cite{Peter}
\begin{equation}
  \bar\beta_2^{\Rsub V} = \beta_2^{\Rsub V} + \frac{\pi^2}{3}\beta_0^3,
\end{equation}
and the minor coefficients $\bar\beta_{0,1}^{\Rsub V}$ are equal to the
scheme-independent values given above. 

To normalize the couplings, we use (\ref{vms}) at ${\bf q}^2 = m_Z^2$.

\subsection{Confining term}

The nonperturbative behaviour of QCD forces between the static heavy quarks  at
long distances $r$ is usually represented by the linear potential (see
discussion in ref.\cite{simon})
\begin{equation}
V^{\rm conf}(r) = k\cdot r, \label{conf}
\end{equation}
which corresponds to the square-law limit for the Wilson loop. 

We can represent this potential in terms of constant chromoelectric field
between the sources posed in the fundamental representation of SU($N_c$). So,
in the Fock-Schwinger gauge of fixed point
$$
x_\mu\cdot A^\mu(x) = 0,
$$
we can represent the gluon field by means of strength tensor \cite{Fort}
$$
A_\mu(x) \approx -\frac{1}{2} x^\nu G_{\nu\mu}(0),
$$
so that for the static quarks separated by the distance $\bf r$
$$
\bar Q_i(0)\; G^a_{m0}(0)\; Q_j(0) = \frac{{\bf r}_m}{r}\; E\; T^a_{ij},
$$
where the heavy quark fields are normalized to unit. Then, the confining
potential is written down as
$$
V^{\rm conf}(r) = \frac{1}{2} g_s\, C_F\, E \cdot r.
$$
Supposing, that the same strength of the field is responsible for the forming
of
gluon condensate, and introducing the colored sources $n_i$, which have to be
averaged in the vacuum, we can easily find \cite{VolLeut}
$$
\langle G^2_{\mu\nu}\rangle = -4\,\langle G^a_{m0}(0) G^a_{m0}(0)\rangle  =
4\,C_F\, E^2 \langle \bar n n\rangle,
$$
where we have supposed the relation
\begin{equation}
\langle \bar n T^a T^b n\rangle = - \langle \bar n T^a n \cdot \bar n
T^b n\rangle,
\label{aver}
\end{equation}
which ensures that the sources conserve the massless of the gluon, and, hence,
the gauge invariance\footnote{The mass term generated by the sources should be
equal to ${\cal L}\sim A_\nu^a A_\nu^b [\bar n T^a T^b n + \bar n T^a n \cdot
\bar n T^b n]$, so that the averaging of sources yields zero, if we suppose
(\ref{aver}).}. Further, it is evident that
$$
\langle \bar n T^a T^b n\rangle = C_F \frac{\delta^{ab}}{N_c^2-1} \langle \bar
n n\rangle .
$$
Then, we conclude that the relation between the strength $E$ and the string
tension depends on the normalization of vacuum sources $n_i$. We put
$$
\langle \bar n_i n_j\rangle = n_l \delta_{ij},
$$
where $n_l$ denotes the number of light \underline{stochastic} flavors, which
is the free parameter of such the representation. Of
course, the value of $n_l$ should be finite even in the case of pure
gluodynamics with no light quarks in the infrared region. Moreover, the light
quark loops could cause the breaking of gluon string, i.e. the strong decays of
higher excitations. We assume that $n_l$ is basically determined by the gluon
dynamics (i.e. the number of colors), and it slightly correlates with the
number of quark flavors. After a simple consideration of potential strength
between two colored sources in the fundamental and adjoint representations,
i.e. the color factors in front of single gluon coulomb potential, we assume
that in the pure gluodynamics the number of stochastic sources substituting for
the vacuum gluons can be accepted in the form\footnote{This assumption
corresponds to the definition of vacuum properties in QCD in terms of notations
under the consideration, which is in agreement with the value of gluon
condensate and Regge trajectories slope.}
$$
n_l = \frac{1}{N_c}\,\frac{C_A}{C_F} = \frac{3}{4} = \frac{1}{4} \tilde n_l, 
$$
where the factor $1/N_c$ normalizes the source to unit, and $C_A/C_F$ is the
appropriate ratio of color charges. To the moment, the shift of $n_l$ in QCD
with light quarks is not explicitly fixed, while the lattice calculations shown
that the dependence of string tension on the number of light quarks is weak
\cite{latstring}. Finally, we find for the linear term of the potential
\begin{equation}
k = \frac{\pi}{\sqrt{C_F N_c \tilde n_l}}\, C_F \sqrt{\langle
\frac{\alpha_s}{\pi} G^2_{\mu\nu}\rangle} = \frac{\pi}{2 \sqrt{N_c}}\, C_F
\sqrt{\langle \frac{\alpha_s}{\pi} G^2_{\mu\nu}\rangle}.
\label{cond}
\end{equation}
The $k$ term is usually represented through a parameter $\alpha^\prime_P$ as
$$
k = \frac{1}{2 \pi \alpha^\prime_P}.
$$
Buchm\"uller and Tye put $\alpha^\prime_P= 1.04$ GeV$^{-2}$, which we use
throughout of this paper. This value of tension, that is related with a slope
of Regge trajectories, can be compared with the estimate following from
(\ref{cond}). At $\langle \frac{\alpha_s}{\pi}
G^2_{\mu\nu}\rangle = (1.6\pm 0.1) \cdot 10^{-2}$ GeV$^4$ \cite{SVZ} we have
found
$$
\alpha^\prime_P = 1.04 \pm 0.03\;\; {\rm GeV}^{-2},
$$
which is in a good agreement with the fixed value\footnote{The ambiguity in the
choice of $n_l$ can change the appropriate value of gluon condensate.}.

The form of (\ref{conf}) corresponds to the limit, when at low virtualities
${\bf q}^2\to 0$ the coupling $\alpha_{\Rsub V}$ tends to
$$
\alpha_{\Rsub V}({\bf q^2}) \to \frac{K}{\bf q^2},
$$
so that 
\begin{equation}
\frac{d \alpha_{\Rsub V}({\bf q^2})}{\ln \bf q^2} \to - \alpha_{\Rsub V}({\bf
q^2}),
\label{lim-c}
\end{equation}
which gives the confinement asymptotics for the $\beta_{\Rsub V}$ function.

A special comment should be done on the role of linear term in the
potential. Considering the power corrections, which can be attributed to
various sources such as the renormalon, topological effects caused by monopoles
and vorteces, deviations from the operator product expansion, the authors of
\cite{zak} argued that this term responsible for the quark confinement can
contribute at short distances, too. This conclusion is essentially different
from the point of view based on the notion about a low energy phase transition
leading to the condensation of gluons and quarks. This condensation provides
the forming of chromoelectric string between the static quarks. Thus, at short
distances (or high virtualities $q^2$) one could expect the decomposition of
condensates, that means the scale of confinement (or the string tension) should
disappear from the physical quantities at large $q^2$. In contrast, the
nonperturbative scale can contribute as the factor in front of power
corrections $1/q^2$ even at $q^2\to \infty$. So, in \cite{zak} several
indications of linear term contribution at small distances were considered. We
repeat the items relevant to the question on the static potential here.

First, the lattice simulation \cite{bali2} does not show any change in the
slope of the full $Q\bar{Q}$ potential as the distances are changed from the
largest to the smallest ones where the coulombic part becomes dominant. Hence,
no rapid energy jump, characteristic for the phase transition, is found on the
lattice. An explicit subtraction of the perturbative corrections at small
distances from the potential in the lattice gluodynamics was performed in
\cite{bali3}. This procedure gives an essential nonzero linear term at very
small distances.

Second, there are the lattice measurements \cite{Fi98} of the fine splitting in
the heavy quarkonium levels as a function of the heavy quark mass. The approach
by Voloshin and Leutwyler \cite{VolLeut} predicts a particular pattern of such
the dependence. Indeed, the multipole expansion of heavy quarkonium interaction
with the external gluon field leads to the dominant contribution by the second
order of chromoelectric dipole. Therefore, the quark distance squared appears
as the leading term in the perturbation due to soft gluons at short distances.
These predictions are very different from the evaluations based on the static
quark potential with the linear term, such as the potential by  Buchm\"uller
and Tye \cite{BT}. The numerical results from the lattice simulations favor the
linear correction to the potential at short distances.

Third, an interesting manifestation of short strings might be the power
corrections to current correlation functions $\Pi_j(q^2)$. Calculations of a
relevant coefficient in front of the $1/q^2$ terms involve the model
assumptions. So, in \cite{CNZ99} it was suggested to simulate this power
correction by a tachyonic gluon mass. The tachyonic mass can imitate the
stringy piece in the potential at short distances. Rather unexpectedly, the use
of the tachyonic gluon mass ($m^2_g = -0.5 \mbox{ GeV}^2$) explains well the
behavior of $\Pi_j(q^2)$ in various channels. This fact implies again we see
the confirmation of short distance linear term in the potential.

Thus, we do not involve any additional assumptions on the possible scale and
properties of quark-gluon condensate decomposition at short distances in the
description of static potential in QCD.

\subsection{Unified $\beta$ function and potential}

Buchm\"uller and Tye supposed the following procedure for the reconstruction of
$\beta$ function in the whole region of charge variation by the known limits of
asymptotic freedom to a given order in $\alpha_s$ and confinement regime. So,
in the framework of asymptotic perturbative theory (PT) to one loop, the
$\beta_{\rm PT}$ is transformed to the Richardson one,
\begin{equation}
\displaystyle
\frac{1}{\beta_{\rm PT}({\EuFrak a})} = -\frac{1}{\beta_0 {\EuFrak a}^2}
\Longrightarrow 
\frac{1}{\beta_{\rm Rich}({\EuFrak a})} = -\frac{1}{\beta_0 {\EuFrak a}^2
\left(1-
\exp\left[-\frac{1}{\beta_0 {\EuFrak a}}\right]\right)}.
\label{Rich}
\end{equation}
The Richardson function has the essential peculiarity at ${\EuFrak a}\to 0$, so
that the expansion is the asymptotic series in ${\EuFrak a}$. At ${\EuFrak
a}\to \infty$ the $\beta$ function tends to the confinement limit represented
in (\ref{lim-c}).

To the two loop accuracy, following in the same way results in the $\beta$
function by Buchm\"uller--Tye,
\begin{equation}
\displaystyle
\frac{1}{\beta_{\rm PT}({\EuFrak a})} = -\frac{1}{\beta_0 {\EuFrak a}^2} +
\frac{\beta_1}{\beta_0^2 {\EuFrak a}} \Longrightarrow 
\frac{1}{\beta_{\rm BT}({\EuFrak a})} = -\frac{1}{\beta_0 {\EuFrak a}^2
\left(1- \exp\left[-\frac{1}{\beta_0 {\EuFrak a}}\right]\right)}+
\frac{\beta_1}{\beta_0^2 {\EuFrak a}}
\exp[-l {\EuFrak a}].
\label{but}
\end{equation}
The exponential factor in the second term contributes to the next order in
${\EuFrak a}$ at small ${\EuFrak a}$, so that the perturbative limit is
restored. However, we can easily find that third coefficient of $\beta_{\rm
BT}$ function is equal to
$$
\beta_{2,\rm BT} = \frac{\beta_1}{\beta_0}
(\beta_1- l \beta_0),
$$
and it is negative at the chosen value of $l=24$ \cite{BT}, which is in
contradiction with the recent result \cite{Peter,Schroed}, shown in
(\ref{b2v}). 

To incorporate the three loop results on the perturbative $\beta$ function, we
introduce
\begin{eqnarray}
\displaystyle
\frac{1}{\beta_{\rm PT}({\EuFrak a})} &=& -\frac{1}{\beta_0 {\EuFrak a}^2} +
\frac{\beta_1+\left(\beta_2^{\Rsub V} - \frac{\beta_1^2}{\beta_0}\right)
{\EuFrak a}}{\beta_0^2 {\EuFrak a}} \Longrightarrow \nonumber \\
\frac{1}{\beta({\EuFrak a})} &=& -\frac{1}{\beta_0 {\EuFrak a}^2 \left(1-
\exp\left[-\frac{1}{\beta_0 {\EuFrak a}}\right]\right)}+
\frac{\beta_1+\left(\beta_2^{\Rsub V} - \frac{\beta_1^2}{\beta_0}\right)
{\EuFrak a}}{\beta_0^2 {\EuFrak a}}
\exp\left[-\frac{l^2 {\EuFrak a}^2}{2}\right],
\label{KKO}
\end{eqnarray}
where again the exponential factor in the second term contributes to the next
order in ${\EuFrak a}\to 0$. In the perturbative limit the usual solution
\begin{eqnarray}
{\EuFrak a}(\mu^2) = \frac{1}{\beta_0 \ln
\frac{\mu^2}{\Lambda^2}}&&\left[1 - 
\frac{\beta_1}{\beta_0^2}\frac{1}{\ln
\frac{\mu^2}{\Lambda^2}} 
\ln \ln \frac{\mu^2}{\Lambda^2} + \right.
\left.
\frac{\beta_1^2}{\beta_0^4}\frac{1}{\ln^2
\frac{\mu^2}{\Lambda^2}} 
\left( \ln^2 \ln \frac{\mu^2}{\Lambda^2} - \ln \ln \frac{\mu^2}{\Lambda^2} -1 +
\frac{\beta_2^{\Rsub V} \beta_0}{\beta_1^2}\right)\right],
\label{3pt}
\end{eqnarray}
is valid. Using the asymptotic limits of (\ref{lim-c}) and (\ref{3pt}), one can
get the equations for any $\beta$ function, satisfying these boundary
conditions, as follows:
\begin{eqnarray}
\ln \frac{\mu^2}{\Lambda^2} & = & \frac{1}{\beta_0 {\EuFrak
a}(\mu^2)}+
\frac{\beta_1}{\beta_0^2} \ln \beta_0 {\EuFrak
a}(\mu^2)+\int_0^{{\EuFrak a}(\mu^2)} dx 
\left[\frac{1}{\beta_0 x^2}-\frac{\beta_1}{\beta_0^2
x}+\frac{1}{\beta(x)}\right], \label{ptlim} \\
\ln \frac{K}{\mu^2} & = & \ln {\EuFrak a}(\mu^2)+\int_{{\EuFrak
a}(\mu^2)}^\infty dx 
\left[\frac{1}{x}+\frac{1}{\beta(x)}\right]. \label{conlim}
\end{eqnarray}
In general, at a given $\beta$ function, Eqs.(\ref{ptlim}) and (\ref{conlim})
determine the connection between the scale $\Lambda$ and the parameter of
linear
potential $K$,
$$
k = 2 \pi C_F K.
$$
Supposing (\ref{KKO}) we can easily integrate out (\ref{ptlim}) to get
the implicit solution of charge dependence on the scale
\begin{eqnarray}
\ln \frac{\mu^2}{\Lambda^2} =&& \ln\left[\exp\left(\frac{1}{\beta_0
{\EuFrak a}(\mu^2)}\right) - 1\right] + \frac{\beta_1}{\beta_0^2}
\left[\ln\frac{\beta_0\sqrt{2}}{l} - \frac{1}{2} \left(\gamma_E +
{\rm E}_1\left[\frac{l^2 {\EuFrak a}^2(\mu^2)}{2}\right]\right)\right]+
\nonumber\\
&&
\frac{\beta_2^{\Rsub V}\beta_0 -\beta_1^2}{\beta_0^3}\,
\frac{\sqrt{\frac{\pi}{2}}}{l}\, {\rm Erf} \left[\frac{l\,
{\EuFrak a}(\mu^2)}{\sqrt{2}}\right],
\label{charge}
\end{eqnarray}
where E$_1[x]=\int_x^\infty dt\, t^{-1} \exp[-t]$ is the exponential integral,
and Erf$[x]=\frac{2}{\sqrt{\pi}}\int_0^x dt\, \exp[-t^2]$ is the error
function.

Eq.(\ref{charge}) can be inverted by the iteration procedure as it was explored
in the derivation of (\ref{3pt}). So, the approximate solution of
(\ref{charge}) has the following form:
\begin{equation}
{\EuFrak a}(\mu^2) = \frac{1}{\beta_0 \ln\left(1+\eta(\mu^2)
\frac{\mu^2}{\Lambda^2}\right)},
\label{eff}
\end{equation}
where
\begin{eqnarray}
\eta(\mu^2) &=& \left(\frac{l}{\beta_0 \sqrt{2}}\right)^
{\frac{\beta_1}{\beta_0^2}} 
\exp\left[
\frac{\beta_1}{2 \beta_0^2} \left(\gamma_E + {\rm
E}_1\left[\frac{l^2
{\EuFrak a}^2_1(\mu^2)}{2}\right]\right)-\frac{\beta_2^{\Rsub
V}\beta_0
-\beta_1^2}{\beta_0^3}\,
\frac{\sqrt{\frac{\pi}{2}}}{l}\, {\rm Erf} \left[\frac{l\,
{\EuFrak a}_1(\mu^2)}{\sqrt{2}}\right]\right],
\end{eqnarray}
while ${\EuFrak a}_1$ is obtained in two iterations
\begin{eqnarray}
\displaystyle
{\EuFrak a}_1(\mu^2) &=& \frac{1}{\beta_0 \ln\left(1+\eta_1(\mu^2)
\frac{\mu^2}{\Lambda^2}\right)}, \\
\eta_1(\mu^2) &=& \left(\frac{l}{\beta_0 \sqrt{2}}\right)^
{\frac{\beta_1}{\beta_0^2}} 
\exp\left[
\frac{\beta_1}{2 \beta_0^2} \left(\gamma_E + {\rm
E}_1\left[\frac{l^2
{\EuFrak a}^2_0(\mu^2)}{2}\right]\right)\right], \\
{\EuFrak a}_0(\mu^2) &=& \frac{1}{\beta_0
\ln\left(1+\frac{\mu^2}{\Lambda^2}\right)}.
\end{eqnarray}
Taking the limit of $\mu^2\to 0$ we find the relation 
\begin{equation}
\ln 4 \pi^2 C_F \alpha^\prime_P \Lambda^2 = \ln \beta_0 +
\frac{\beta_1}{2
\beta_0^2} \left(\gamma_E + \frac{l^2}{2\beta_0^2}\right)-\frac{\beta_2^{\Rsub
V}\beta_0 -\beta_1^2}{\beta_0^3}\,
\frac{\sqrt{\frac{\pi}{2}}}{l},
\end{equation}
which completely fixes the $\beta$ function and charge in terms of scale
$\Lambda$ and the slope $\alpha^\prime_P$, since we have expressed the
parameter $l$ in terms of above quantities.

Remember, that at $\mu^2\to\infty$ the perturbative expression (\ref{3pt})
becomes valid as the limit of effective charge (\ref{eff}).

To the moment we are ready to discuss the numerical values of parameters.

\subsection{Setting the scales}
As we have already mentioned the slope of Regge trajectories, determining the
linear part of potential, is fixed as
$$
\alpha^\prime_P = 1.04\;{\rm GeV}^{-2}.
$$
We use also the measured value of QCD coupling constant \cite{PDG} and pose
$$
\alpha_s^{\overline{\Rsub MS}}(m_Z^2) = 0.123,
$$
as the basic input of the potential.

At the given choice of normalization value for the QCD coupling constant we
get the scale $\Lambda^{\overline{\Rsub MS}}_{n_f=5}\approx 273$ MeV, which
certainly differs from the world average value resulted in the analysis of PDG
\cite{PDG}, where $\Lambda^{\overline{\Rsub MS}}_{n_f=5}\approx
208^{+25}_{-23}$ MeV, that corresponds to the coupling constant
$\alpha_s^{\overline{\Rsub MS}}(m_Z^2) = 0.1181\pm 0.002$ \cite{PDG}. However,
this average value including various data is generally determined by the most
precise measurements: the data on the hadronic events in the peak of $Z$ boson
at LEP (the hadronic width), the decays of $\tau$ lepton, the data on the deep
inelastic scattering (DIS) for leptons off nucleons and the lattice simulations
for the systems of heavy quarkonia. In this set of estimates, the high energy
measurements at LEP for $Z$ and at HERA for the evolution of nucleon structure
functions give the average values $\alpha_s^{\overline{\Rsub MS}}(m_Z^2) =
0.123\pm 0.004$ and $\alpha_s^{\overline{\Rsub MS}}(m_Z^2) = 0.122\pm 0.004$,
respectively, while the evolution of structure functions at low virtualities,
where an ambiguity in the description of nonperturbative effects and
contributions of higher twists are essential, as well as the energy-dependent
sum rules for the structure functions at low energies significantly displace
down the common average value for the coupling constant extracted from the DIS
data. Thus, we argue that the methodical uncertainty for such the averaging is
underestimated, since the low-energy data have got some little calculated
sources of theoretical uncertainties. The analysis of data on the decays of
$\tau$ lepton resulting in $\alpha_s^{\overline{\Rsub MS}}(m_Z^2) = 0.121\pm
0.003$, is based on the sum rules, where the control of nonperturbative
corrections is much better than in DIS, though there are some theoretical
problems on the formulation of sum rules in the region of physical states in
contrast with the classic variant of sum rules in the deep euclidean region.
Finally, the lattice simulations investigate the splitting between the states
of heavy quarkonia, i.e. they operate with the low-energy data and rely on the
approximation with the zero number of light quarks $n_f=0$ or $n_f=2$ under the
extrapolation to both the real number of $n_f=3$ and the region of high
virtualities due to the evolution. A high accuracy of such lattice estimates is
announced. As we have seen the spectroscopic characteristics for the systems of
heavy quarks need an extremely careful interpretation, since the evolution of
potential parameters from the region of bound states to the high virtualities
is affected by the nonperturbative factors. Thus, we see that the normalization
value of QCD coupling constant accepted above agrees with the direct
high-energy measurements, while the data obtained at low energies allow the
agreement, if we take into account their systematic uncertainties, which are
not well estimated.

Note that the decrease of normalization value to $\alpha_s^{\overline{\Rsub
MS}}(m_Z^2) = 0.120$, for example, leads to the discrepancy with the data on
the splitting of heavy quarkonium masses between the levels of $1S$ and $2S$
states, which is very sensitive to the normalization of QCD coupling constant,
so that instead of $M(2S)-M(1S)\approx 580$ MeV we get the value which is less
by about $100$ MeV. In this respect, the variation of other dimensional
parameter, the Regge trajectory slope, from the accepted value of
$\alpha^{\prime}_P=1.04$ GeV$^{-2}$ to $\alpha^{\prime}_P=0.87$ GeV$^{-2}$
leads to unessential change in both the splitting and the corresponding value
for the scale in the coupling constant evolved to low virtualities.

Then, we evaluate
$$
\alpha_{{\Rsub V}}(m_Z^2) \approx 0.1306,
$$
and put it as the normalization point for ${\EuFrak a}(m_Z^2)=\alpha_{{\Rsub
V}}(m_Z^2)/(4\pi)$. Further, we find the following values of $\Lambda$ for the
effective charge, depending on the number of active flavors:
\begin{eqnarray}
\Lambda_{n_f=3} &=& 643.48\;{\rm MeV,} \;\;\; l=56,\\
\Lambda_{n_f=4} &=& 495.24\;{\rm MeV,} \;\;\; l=37.876,\\
\Lambda_{n_f=5} &=& 369.99\;{\rm MeV,} \;\;\; l=23.8967,
\end{eqnarray}
where we set the threshold values for the switching the number of flavors to be
equal to $m_5 = 4.6$ GeV and $m_4=1.5$ GeV. After such the fixing the momentum
space dependence of the charge, we perform the Fourier transform to get 
\begin{equation}
V(r) = k \cdot r - \frac{8 C_F}{r} u(r),
\label{r<}
\end{equation}
with
$$
u(r) = \int_0^\infty \frac{d q}{q}\, \left({\EuFrak
a}(q^2)-\frac{K}{q^2}\right) \sin
(q\cdot r),
$$
which is calculated numerically at $r>0.01$ fm and represented in the
MATHEMATICA file in the format of notebook at the site {\sf
http://www.ihep.su/$^\sim$kiselev/Potential.nb}.

Note, that at short distances the potential behaviour is purely perturbative,
so
that at $r<0.01$ fm we put 
\begin{equation}
V(r) = - C_F\, \frac{\bar \alpha_{{\Rsub V}}(1/r^2)}{r},
\label{r>}
\end{equation}
where the running $\bar\alpha_{{\Rsub V}}(1/r^2)$ is given by eq.(\ref{3pt})
with the appropriate value of $\bar \beta_2^{\Rsub V}$ at $n_f = 5$, and with
the matching with the potential (\ref{r<}) at $r_s=0.01$ fm, where we have
found
$$
\bar\alpha_{\Rsub V}(1/r_s^2) = 0.22213,
$$
which implies $\Lambda_{n_f=5}^{\overline{\Rsub V}}=617.42$ MeV.

\begin{figure}[ph]
\setlength{\unitlength}{1mm}
\begin{center}
\begin{picture}(100,90)
\put(5,5){\epsfxsize=9cm \epsfbox{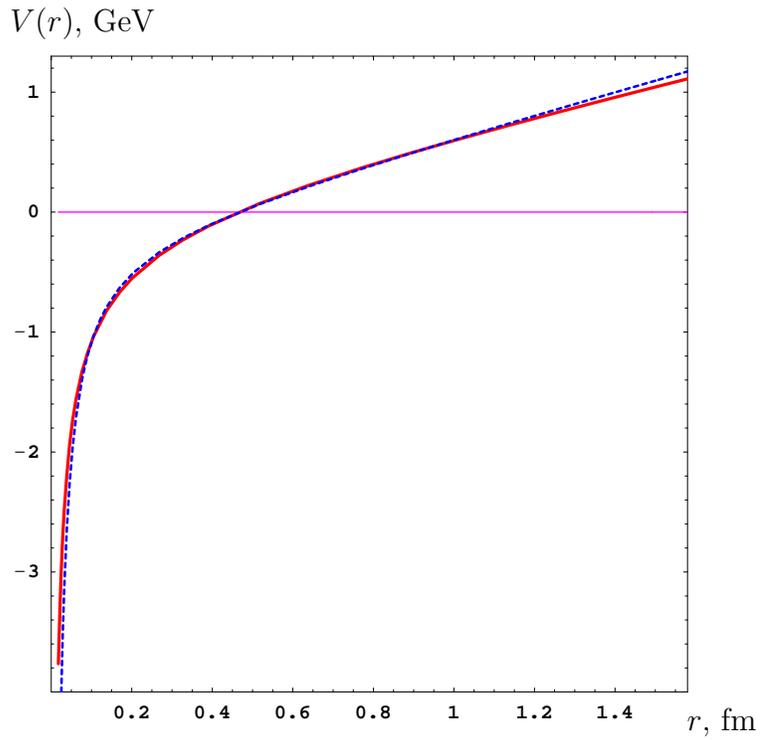}}
\put(95,5){$r$, fm}
\put(5,98){$V(r)$, GeV}
\end{picture}
\end{center}

\caption{The potential of static heavy quarks in QCD (solid line) in comparison
with the Cornell model (dashed line) 
(up to an additive shift of energy scale).} 
\label{pot-fig}
\end{figure}

Thus, we have completely determined the model for the potential of static heavy
quarks in QCD. In Fig. \ref{pot-fig} we present it versus the distance between
the quarks. As we can see the potential is very close to what was obtained in
the Cornell model in the phenomenological manner by fitting the mass spectra of
heavy quarkonia.

\begin{figure}[ph]
\setlength{\unitlength}{1mm}
\begin{center}
\begin{picture}(100,100)
\put(5,5){\epsfxsize=9cm \epsfbox{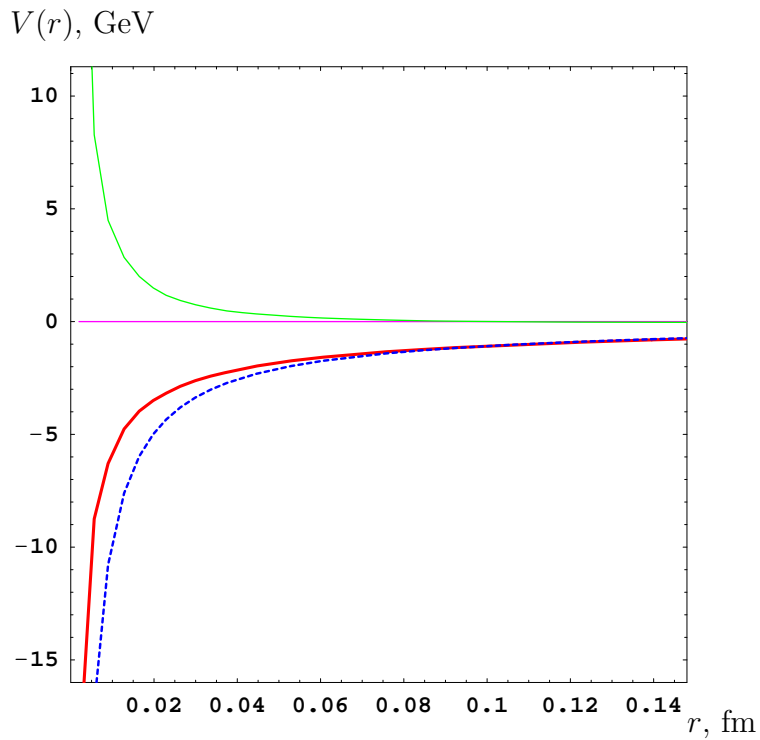}}
\put(95,5){$r$, fm}
\put(5,98){$V(r)$, GeV}
\end{picture}
\end{center}

\caption{The potential of static heavy quarks in QCD (solid line) in comparison
with the Cornell model (dashed line, up to an additive shift of energy scale)
and the difference between them (upper curve) at short distances as caused by
the running of coupling in QCD.} 
\label{diffCornell}
\end{figure}

The visual deviation between the QCD potential derived and the Cornell model at
long distances is caused by a numerical difference in the choice of string
tension: we adopt the value given by Buchm\"uller and Tye, while in the Cornell
model the tension is slightly greater than that of we have used. A more
essential point is the deviation between the potentials at short distances (see
Fig. \ref{diffCornell}), because of clear physical reason, the running of
coupling constant in QCD in contrast to the constant effective value in the
Cornell model.

\begin{figure}[th]
\setlength{\unitlength}{1mm}
\begin{center}
\begin{picture}(100,90)
\put(5,5){\epsfxsize=9cm \epsfbox{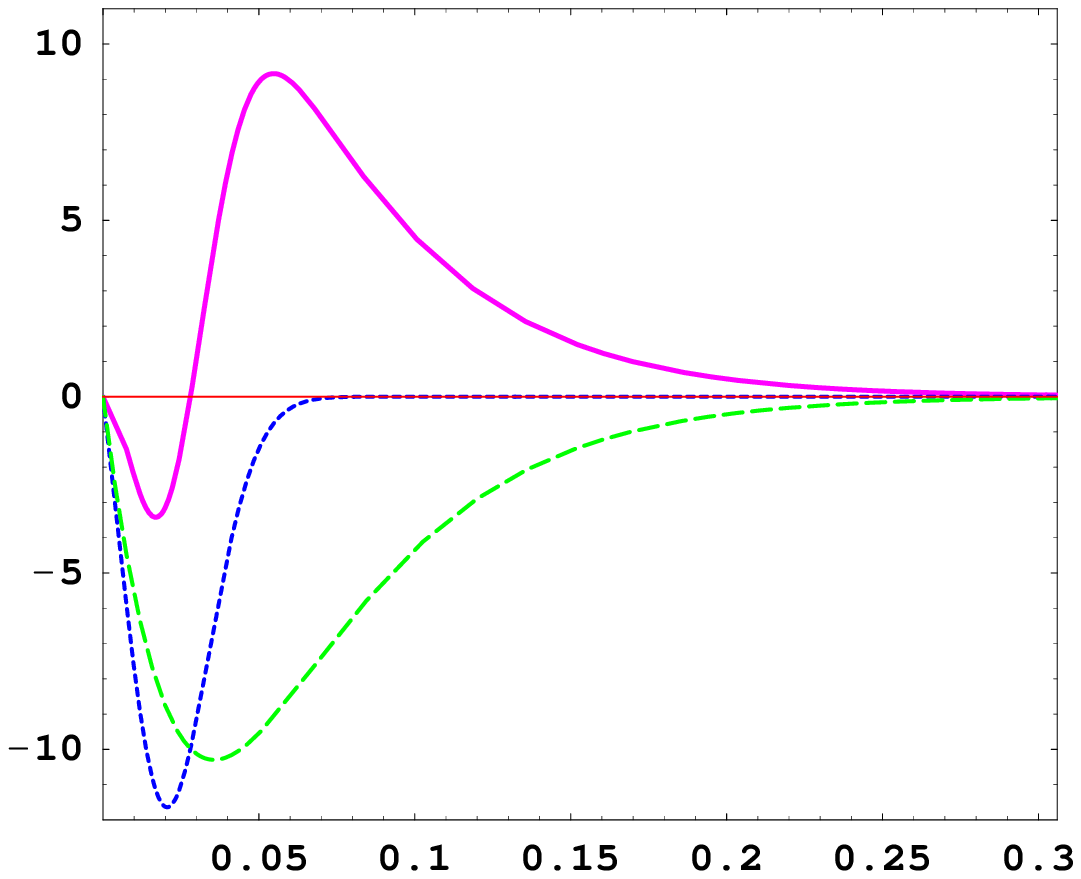}}
\put(99,5){$\EuFrak a$}
\put(5,85){$\displaystyle \frac{\Delta \beta}{|\beta|}$, \%}
\end{picture}
\end{center}

\caption{The differences between the $\beta$ functions versus the effective
charge. The value of $(\beta-\beta_{\rm BT})/ |\beta|$ is shown by the solid
line, $(\beta-\beta_{\rm Rich})/ |\beta|$ is given by the short-dashed line,
and $(\beta_{\rm BT}-\beta_{\rm Rich})/ |\beta_{\rm BT}|$ is represented by the
long-dashed line.} 
\label{dif}
\end{figure}

To compare, we show the differences between the $\beta$ functions (\ref{Rich}),
(\ref{but}) and (\ref{KKO}) in Fig. \ref{dif} at the fixed values of $l$ and
$n_f=3$. Wee see that the asymptotic perturbative expansion of $\beta$ at
${\EuFrak a}\to 0$ dominates at ${\EuFrak a} < {\EuFrak a}_0$, where ${\EuFrak
a}_0 \approx 0.03$ corresponding to $\alpha_{{\Rsub V},0} \approx 0.37$. This
value of coupling $\alpha_{{\Rsub V},0}$ coincides with the effective coulomb
constant used in the Cornell model. At larger values of coupling the
contributions related with the confinement regime are essential.

Two comments are to the point. First, the resulting potential is obtained by
the perturbative normalization to the measured value of
$\alpha_s^{\overline{\Rsub MS}}(m_Z^2)$ as combined with the three-loop
evolution to the lower virtualities. Second, the running of coupling constant
is modified (numerically the deviation from the perturbative regime begins at
$\mu < 3-4$ GeV) to reach the confinement limit at $\mu\to 0$, so that the
perturbative connection between the scales $\Lambda$ and
$\Lambda^{\overline{\Rsub MS}}$ is broken at virtualities under touch by the
charmed and bottom quarks, that was the reason for the error in the assignment
of $\Lambda^{\overline{\Rsub MS}}$ by Buchm\"uller and Tye.

\section{Heavy quark masses and leptonic constants}
Considering the characteristics of heavy quark bound states we should emphasize
a significant necessity to certainly separate two distinct theoretical
problems. The first problem is the calculation of heavy quark potential, where
the leading approximation is the static limit of $m_Q\to \infty$ in the
operator product expansion over the powers of inverse heavy quark mass. we have
considered this problem  in the previous section. The other problem is the
calculation of bound state masses. In the heavy quarkonium the kinetic energy
of quark motion is comparable with the potential energy. So, the leading
approximation for the effective lagrangian in the operator product expansion
over the inverse heavy quark mass is the sum of nonrelativistic kinetic term
and the static potential, which give the dominant contribution in the
Schr\"odinger equation for the bound states. Corrections are relativistic terms
in the kinetic energy and perturbations of the static potential in the form of
operators suppressed by the inverse powers of heavy quark mass, as well as
nonpotential retardation effects. The magnitude of such the corrections can be
restricted numerically, that leads to a systematic uncertainty in the
calculations of mass spectra for the heavy quarkonia in the framework of
potential approach with the static potential.

\subsection{Masses}
The determination of potential provides us with the extraction of heavy quark
masses in the static approximation by comparison of heavy quarkonium
mass-spectra with the calculated ones. The predicted charmonium and bottomonium
masses are presented in Tables\footnote{We suppose that the $\psi(3770)$-state
is a mixture of $3S$ and $3D$ levels with unessential shift of $3D$-mass.}
\ref{tc} and \ref{tb} at the following values of heavy quark masses in the
potential approach 
\begin{equation}
m_c^{\Rsub V} = 1.468\;{\rm GeV,}\;\;\;
m_b^{\Rsub V} = 4.873\;{\rm GeV,}
\label{mcmb}
\end{equation}
with no taking into account relativistic corrections, which can be sizable for
the charmonium (say, $\Delta M(\bar c c) \sim 40$ MeV). To the moment, the only
measured splitting of $nS$-levels is that of $\eta_c$ and $J/\psi$, which
allows us to evaluate the so-called spin-averaged mass
$$
\overline{M}(1S) = (3 M_{J/\psi}+M_{\eta_c})/4.
$$ 
Supposing the simple relation \cite{pmQR}: $\overline{M}(ns) =
M_V(nS)-\frac{1}{4 n}(M_{J/\psi}-M_{\eta_c})$, we estimate also the
expected values for the excited states with an accuracy better than 10 MeV, we
think. For the $P$-wave levels we explore the masses
$$
\overline{M}(P) = M_1+\frac{1}{3}(M_2-M_0)+\frac{2}{9}(M_2-M_1+2(M_0-M_1)),
$$
where $M_J$ denotes the mass of state with the total spin $J$ and the sum of
quark spins $S=1$, and we have supposed the spin-dependent forces in the form
$$
V_{SD} = A ({\bf L\cdot S}) + B ({\bf L\cdot S})^2 -\frac{1}{3} B {\bf
L}^2\cdot {\bf S}^2,
$$ 
where the third term corresponds to the third term in the above expression for
$\overline{M}(P)$ 
\begin{table}[th]
\caption{The masses of charmonium as predicted in the present paper (K$^2$O) in
comparison with the experimental data elaborated as described in the text.}
\begin{center}
\begin{tabular}{|l|c|c||l|c|c|}
state, $nL$ & $M$, K$^2$O & $\overline{M}$, exp. & state, $nL$ & $M$, K$^2$O &
$\overline{M}$, exp. \\
\hline
1S & 3.068 & 3.068 & 2P & 3.493 & 3.525 \\
2S & 3.670 & 3.671 & 3P & 3.941 &  ---  \\
3S & 4.092 & 4.040 & 3D & 3.785 & 3.770 \\
\end{tabular}
\end{center}
\label{tc}
\end{table}
\noindent
and it results in the $L$-dependent shift of levels.

We have supposed also 
$$
\displaystyle
M_{\Upsilon}-M_{\eta_b}\approx
\frac{\alpha_s(m_b)}{\alpha_s(m_c)} \frac{m_c^2}{m_b^2}\frac{|R_{\bar b
b}(0)|^2}{|R_{\bar c c}(0)|^2} (M_{J/\psi}-M_{\eta_c})\approx 56\; {\rm MeV.}
$$

We have found that the sizes of quarkonia are the same as they were predicted
by Buchm\"uller and Tye, while the masses of states are slightly different
since we have used the other prescription for the input values of ground state
masses:
$$
M_{\bar c c}(1S) = 3.068\;{\rm GeV,}\;\;\;
M_{\bar b b}(1S) = 9.446\;{\rm GeV.}
$$

Then, we predict the masses of $\bar b c$ quarkonium\footnote{The experimental
error in the ground state mass is still large, $\delta M=\pm 0.39$ GeV
\cite{cdf}.}, as shown in Table \ref{bc}. The calculated values of masses agree
with those of estimated in the Buchm\"uller--Tye and Martin potentials
\cite{eichten}. The wave functions at the origin are related with the
production rates of heavy quarkonia. These parameters are close to what was
predicted in the BT potential, but slightly smaller because of both the change
in the charmed quark mass and the asymptotic behaviour at $r\to 0$.
\begin{table}[th]
\caption{The masses of bottomonium as predicted in the present paper (K$^2$O)
in comparison with the experimental data elaborated as described in the text.}
\begin{center}
\begin{tabular}{|l|r|r||l|r|r|}
state, $nL$ & $M$, K$^2$O & $\overline{M}$, exp. & state, $nL$ & $M$, K$^2$O &
$\overline{M}$, exp. \\
\hline
1S & 9.446  & 9.446  & 2P & 9.879  &  9.900 \\
2S & 10.004 & 10.013 & 3P & 10.239 & 10.260 \\
3S & 10.340 & 10.348 & 3D & 10.132 &  ---  \\
4S & 10.606 & 10.575 & 5S & 10.835 & 10.865 \\
\end{tabular}
\end{center}
\label{tb}
\end{table}
\begin{table}[th]
\caption{The masses of $\bar b c$ as predicted in the present paper (K$^2$O) in
comparison with the experimental data.}
\begin{center}
\begin{tabular}{|l|c|c||l|c|c|}
state, $nL$ & $M$, K$^2$O &  $\overline{M}$, exp. & state, $nL$ &
$M$, K$^2$O &
$\overline{M}$, exp. \\
\hline 
1S & 6.322 & 6.40  & 2P & 6.739  &  --- \\
2S & 6.895 &  ---  & 3P & 7.148  &  ---  \\
3S & 7.279 &  ---  & 3D & 7.013  &  --- \\
\end{tabular}
\end{center}
\label{bc}
\end{table}

\begin{figure}[ph]
\setlength{\unitlength}{1mm}
\begin{center}
\begin{picture}(100,100)
\put(5,5){\epsfxsize=9cm \epsfbox{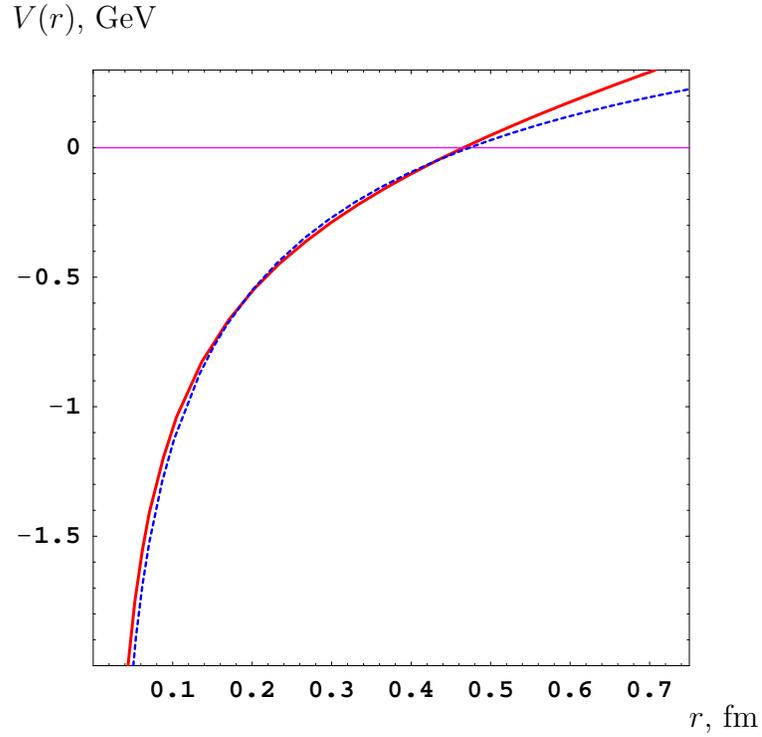}}
\put(95,5){$r$, fm}
\put(5,98){$V(r)$, GeV}
\end{picture}
\end{center}

\caption{The potential of static heavy quarks in QCD (solid line) in comparison
with the perturbative term (\ref{Vr}) at $\mu=1.5$ GeV (dashed line) (up to an
additive shift of energy scale).} 
\label{pot1.5}
\end{figure}

\begin{figure}[ph]
\setlength{\unitlength}{1mm}
\begin{center}
\begin{picture}(100,97)
\put(5,5){\epsfxsize=9cm \epsfbox{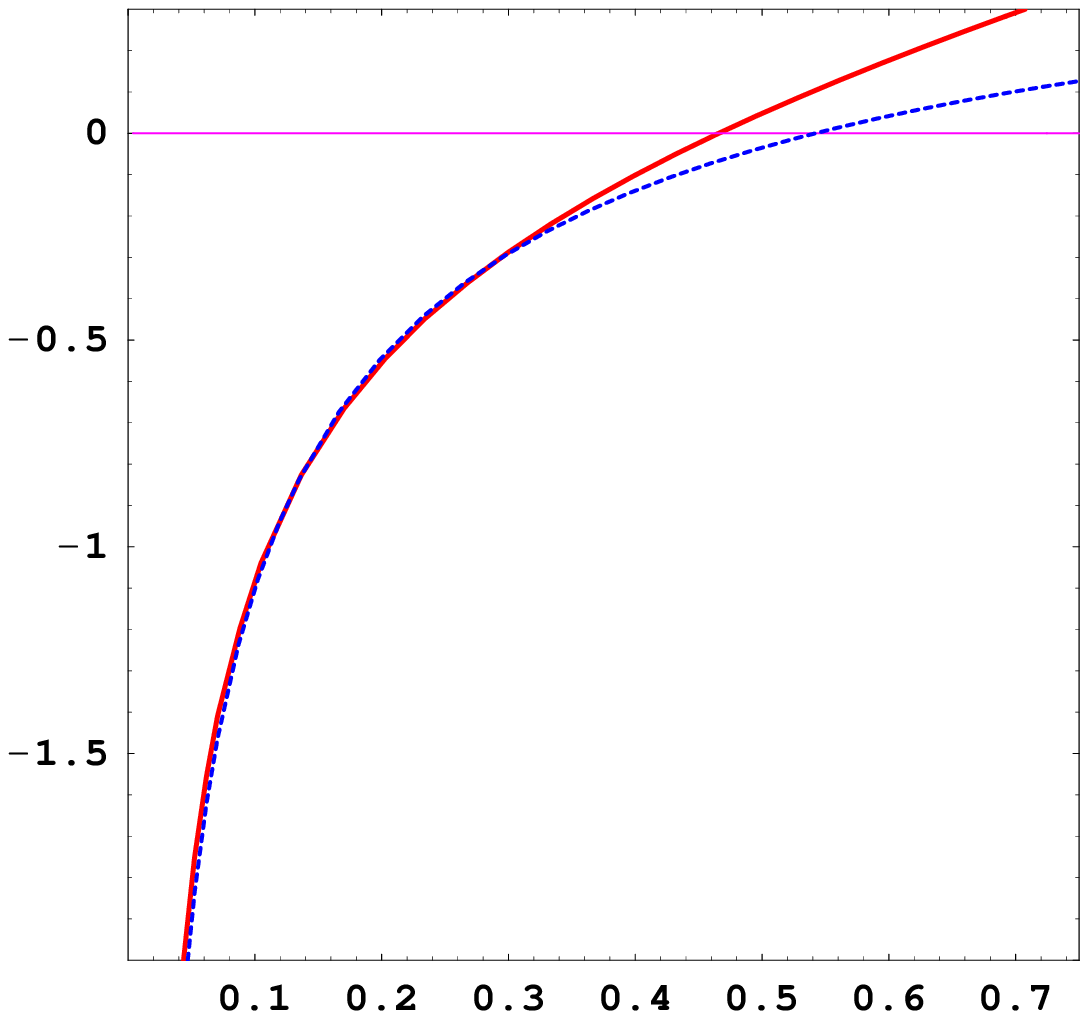}}
\put(95,5){$r$, fm}
\put(5,98){$V(r)$, GeV}
\end{picture}
\end{center}

\caption{The potential of static heavy quarks in QCD (solid line) in comparison
with the perturbative term (\ref{Vr}) at $\mu=2.0$ GeV (dashed line) 
(up to an additive shift of energy scale).} 
\label{pot2.0}
\end{figure}

\begin{figure}[th]
\setlength{\unitlength}{1mm}
\begin{center}
\begin{picture}(100,96)
\put(5,5){\epsfxsize=9cm \epsfbox{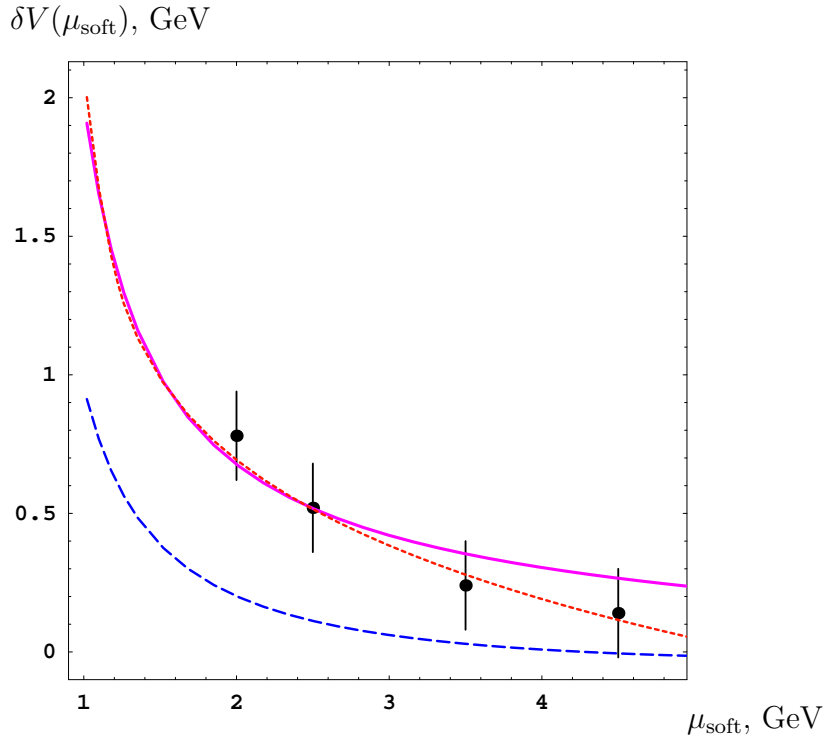}}
\put(95,5){$\mu_{\rm soft}$, GeV}
\put(5,98){$\delta V(\mu_{\rm soft})$, GeV}
\end{picture}
\end{center}

\caption{The value of additive shift of energy scale to match the perturbative
$\mu$-dependent potential with that of calculated in QCD. The solid and dashed
lines correspond to the two and one loop matching. The points give the result
of sum rules for bottomonium in comparison with the dotted curve following from
the relation between the running and pole masses at scale $\mu$.} 
\label{shift}
\end{figure}

To the moment we have fixed the potential masses of heavy quarks (\ref{mcmb})
as independent of scale. To compare with the masses evaluated in the framework
of QCD sum rules, we note that in the sum rules for the heavy quarkonia one
usually explores the NRQCD \cite{NRQCD} with the perturbative potential
(\ref{Vr}) explicitly dependent of the normalization point $\mu$ (referred as
$\mu_{\rm soft}$ in the SR). We have checked that at short distances and high
$\mu_{\rm soft}$ the perturbative potential (\ref{Vr}) and that of present
paper coincide with each other, while a deviation appears at $r\gg 1/\mu_{\rm
soft}$. However, at the distances characteristic for the ground states of heavy
quarkonia: $\langle r_{\bar b b(1S)}\rangle \approx 0.22$ fm and $\langle
r_{\bar c c(1S)}\rangle \approx 0.42$ fm, the shape of the potential can be
approximated by the perturbative term at $\mu_{\rm soft}=1.5-2.0$ GeV (see
Figs. \ref{pot1.5} and \ref{pot2.0}) with the additive shift of energy scale
$\delta V(\mu_{\rm soft})$, which is defined by the expression
\begin{equation}
\delta V(\mu_{\rm soft}) = \left[ V(r) - V_{\Rsub pert}(r;\mu_{\rm
soft})\right] \left|_{_{\scriptstyle r = \frac{1}{\mu_{\rm
soft}}\zeta}}\right.,
\label{def-dV}
\end{equation}
where the parameter $\zeta$ has been put in the region of $\zeta=1-2$, where
the energy shift $\delta V$ has got a little variation about 30-40 MeV, which
is, on the first hand, a characteristic uncertainty of potential approach, and
on the other hand, it points to a similar form of perturbative potential with
the calculated model potential in the region of distance variation. The
dependence of energy shift is represented in Fig. \ref{shift}.

So, if we redefine the heavy quark masses\footnote{This redefinition is the
indication of perturbative renormalon (see review in \cite{mb}).
Indeed, there are two sources for the deviation $\delta V$. The first is the
linear confining term in the potential of static quarks. However, it is a small
fraction of $\delta V$. The second source is the infrared singularity in the
perturbative running coupling. One can easily find that subtracting the
singular term of the form $\sim \frac{1}{\mu_{\rm soft}-\Lambda}$ from $\delta
V$ results in a small value slowly depending on $\mu_{\rm soft}$. In the
effective theory for the nonrelativistic heavy quarks, the subtraction
that connects the pole mass and the threshold mass can be calculated explicitly
(see \cite{Hoangetal} and references therein).} by
$$
m^{\rm pole}(\mu)_{b,\,c} = m_{b,\,c}^{\Rsub V}+ \frac{1}{2} \delta V(\mu),
$$
the solution of Schr\"odinger equation with the perturbative potential and
$m^{\rm pole}(\mu)$ results in the quarkonia masses close to the experimental
values. Thus, we have matched the values of potential masses $m^{\Rsub V}$ in
the QCD potential with the perturbative pole masses standing in the two-loop
calculations. We stress that the dependence on the soft scale in both the
energy shift $\delta V(\mu)$ and the pole mass $m^{\rm pole}(\mu)$ does not
reflect a nonzero anomalous dimension, since these quantities are
renormalization group invariants. This scale dependence is due to the
truncation of perturbative expansion, wherein the coefficients in front of
powers of coupling constant can contain the factorial growth (the renormalon),
so that even at the scale close to the charmed quark mass the infrared
singularity in the running coupling constant of QCD provides the significant
custodial scale dependence.

Numerically, we estimate the running masses $\bar m(\bar m)$ in the
$\overline{\rm MS}$ scheme using the two and three-loop relations with the pole
mass derived in \cite{broad,kmel} and adjusting the scale $\mu_{\rm soft}$ to
be equal to $\bar m$. So, in two loops \cite{broad} we get
$$
\bar m_c(\bar m_c)_{\Rsub 2\, loops} = 1.40\pm 0.09\;{\rm GeV},\;\;\;
\bar m_b(\bar m_b)_{\Rsub 2\, loops} = 4.20\pm 0.06\;{\rm GeV},
$$

\noindent
while the three-loop approximation \cite{kmel}, which is consistent with the
three-loop evolution of coupling constant, results in slightly smaller masses,
especially, for the charmed quark, where the uncertainty of estimate increases
because of stronger sensitivity of quantities involved to the scale variation,
$$
\bar m_c(\bar m_c)_{\Rsub 3\, loops} = 1.17\pm 0.10\;{\rm GeV},\;\;\;
\bar m_b(\bar m_b)_{\Rsub 3\, loops} = 4.15\pm 0.06\;{\rm GeV},
$$
which are in agreement with the various estimates in the sum rules on $m_b$
\cite{pp,ben,mel,hoang} and $m_c$ \cite{paver}\footnote{Note, there is the
difference between the usually quoted values of $\bar m(\bar m)$ and $\bar
m(m^{\rm pole})$.}. 

The uncertainty of estimates is determined by the deviations in the
calculations of heavy quarkonium masses $\frac{1}{2}\delta M =20$ MeV (as shown
in Tables \ref{tc} and \ref{tb}) and the error in the extraction of $\delta V$
mentioned above. The uncertainty in the running mass of charmed quark is
slightly larger than in the bottom mass, since, in addition, its value is more
sensitive to a small variation of scale, pole mass and energy shift.

Note, that the calculations in the framework of sum rules were performed for
the $b$-quark mass in both the full QCD \cite{jamin} and the effective theory
of nonrelativistic heavy quarks NRQCD \cite{pp,ben,mel,hoang,mbv}. The mass
extraction of Ref.\cite{mbv} has been carried out in the nonrelativistic
effective theory at next-to-leading order (NLO), whereas Refs.
\cite{pp,ben,mel,hoang} carried out NNLO analyses in the same framework. The
calculations in the nonrelativistic effective theory are the calculations in
the framework of first principles in QCD, where the results of full QCD are
determined in a systematic expansion in $\alpha_s$ and the velocity. In Ref.
\cite{jamin} the analysis has also been carried out in the nonrelativistic
situation, but no systematic expansion in $\alpha_s$ and the velocity has been
carried out. That the results for the $\overline{\rm MS}$ mass obtained in
\cite{jamin} agree with the other analyses is not understood and requires
further examination (see the conclusions of \cite{Jamin}).

Recently, the charmed quark mass was evaluated from the NRQCD
sum rules in \cite{eide}, so that the result on the running mass is in a good
agreement with the value given above, too. There is a recent sum rule
extraction \cite{Hoanghep} of the bottom $\overline{\rm MS}$ mass, where the
charmed quark mass effects are also included. The estimate of potential
approach under consideration is in a good agreement with this recent SR
result.

In \cite{mel} the dependence of `pole' mass on the scale $\mu_{\rm soft}$ was
explicitly calculated in the N$^2$LO. The uncertainty of mass extraction from
the sum rules for bottomonium was given by $0.1$ GeV for the running
$\overline{\rm MS}$ mass and $0.06$ GeV for the low-energy running mass
(`kinetic' mass). The result on the $b$-quark pole mass depends on both the
scale of calculations and the order in $\alpha_s$ of perturbative QCD. To
compare the results in the sum rules with those of given in the present paper
we fix the order in $\alpha_s$ by the two-loop corrections.  Then we have found
that, say, at $\mu_{\rm soft}=2.5$ GeV the results of estimates in the
perturbative potential approach and in the framework of sum rules are the same
within the uncertainty mentioned. So, putting the above value as the matching
point we show the sum rule results in the form of energy shift in
Fig. \ref{shift}. For the sake of representability in Fig. \ref{shift} we show
the $\mu$-dependent `pole' mass extracted in \cite{mel} with the uncertainty of
$\delta m = 80$ MeV, which is characteristic inherent error for the
short-distance masses in the analysis of \cite{mel}. Despite of various choice
for the normalization of QCD coupling constant (in \cite{mel}
$\alpha_s^{\overline{\Rsub MS}}(m^2_Z)=0.118$), we see a good agreement between
the $\mu$-dependencies of both the energy shift in the perturbative potential
with respect to the static potential of QCD and the variation of perturbative
`pole' mass of $b$-quark in the sum rules of QCD. As for the one-loop matching
of perturbative potential, we mention only that the corresponding sum rules in
the NLO give the value of energy shift close to zero at $\mu_{\rm soft} > 2$
GeV within the uncertainty of the method, and this estimate is consistent with
the result of potential approach as shown in Fig. \ref{shift}. Thus, the energy
shift of perturbative potential with the two-loop matching of $\rm V$ and
$\overline{\rm MS}$ schemes indicates the form of QCD potential in agreement
with the corresponding soft scale dependence of perturbative pole mass in sum
rules of QCD for the bottomonium.

To the moment we can compare the obtained $\mu$-dependence of `pole' mass with
the relation between the running $\overline{\rm MS}$-mass of heavy quark and
the pole mass derived in \cite{taras}, where we find
\begin{equation}
m^{\rm pole} = \bar m(\mu) \left(1+ c_1(\mu) \frac{\alpha_s^{\overline{\Rsub
MS}}(\mu^2)}{4 \pi} +c_2(\mu) \left(\frac{\alpha_s^{\overline{\Rsub
MS}}(\mu^2)}{4 \pi}\right)^2\right),
\label{pole}
\end{equation}
with
\begin{eqnarray}
c_1(\mu) &=&  C_F (4+3 {L}),\\
c_2(\mu) &=&  C_F C_A \left(\frac{1111}{24}- 8 \zeta(2)- 4 I_3(1)+\frac{185}{6}
L+\frac{11}{2}L^2\right)\nonumber\\
&&- C_F T_F n_f \left(\frac{71}{6}+ 8 \zeta(2) + \frac{26}{3} L + 2 L^2
\right)\\
&& + C_F^2 \left(\frac{121}{8}+30 \zeta(2)+8 I_3(1)+\frac{27}{2} L +
\frac{9}{2}L^2\right) - 12 C_F T_F (1-2 \zeta(2)),\nonumber
\end{eqnarray}
where $I_3(1) = \frac{3}{2}\zeta(3)- 6 \zeta(2)\ln 2$, and $L=2\ln(\mu/m^{\rm
pole})$. At $\mu=m^{\rm pole}$, the result of \cite{broad} is reproduced. We
check that the logs in the definitions of $c_{1,2}$ can be removed by the
expression of running values $\bar m(\mu)$ and $\alpha_s^{\overline{\Rsub
MS}}(\mu)$ in terms of $\bar m(m^{\rm pole})$ and $\alpha_s^{\overline{\Rsub
MS}}(m^{\rm pole})$ in (\ref{pole}). Nevertheless, we find that the explicit
$\mu$-dependence in (\ref{pole}) repeats the form of renormalon contribution as
we see it in the perturbative potential, where the similar effect takes place
because of both the truncation of perturbative series and the infrared pole in
the running coupling constant of QCD. Following (\ref{pole}), we show the value
of difference $2(m^{\rm pole}_b(\mu)-m_b)$ in Fig. \ref{shift} at $\bar m(\bar
m) =4.3$ GeV. We see that, first, the results of QCD sum rules in \cite{mel}
agree with the values expected from (\ref{pole}), and second, the
$\mu$-dependent shift of pole mass approximately coincides with the shift of
perturbative potential with respect to the static QCD potential free off
renormalon ambiguity caused by infrared singularity of perturbative coupling
constant at finite energy scale. This fact implies the cancellation of infrared
uncertainties. Thus, we can define the unambiguous pole mass by
\begin{equation}
\hat m^{\rm pole} = m^{\rm pole}(\mu) - \frac{1}{2}\delta V(\mu),
\label{redef}
\end{equation}
where we use the pole mass of (\ref{pole}). The basis for the validity of
(\ref{redef}) was observed in \cite{cancel}, where in the
context of perturbative bottom mass extractions, the cancellation of the
leading renormalon at $u=1/2$ of the Borel plane in the total static
perturbative energy of a heavy $Q\bar Q$ pair was shown.

We find that for the bottom quark
the defined mass is given by the value of mass extracted from the potential
approach
$$
\hat m_b^{\rm pole} \approx m_b^{\Rsub V},
$$
with the accuracy about $80$ MeV.

\subsection{Heavy quark masses and pNRQCD}
In this section we discuss the modern development in the theory of heavy
quarkonium $QQ'$ on the basis of effective theory called pNRQCD \cite{pNRQCD},
naturally incorporating the potential interactions between the heavy quarks and
external ultrasoft fields in QCD, and compare the pNRQCD results with the
values of heavy quark masses obtained above in the QCD potential of static
quarks.

First, pNRQCD argues that in the heavy quarkonium the nonrelativistic motion of
heavy quarks inside the bound state allows us to introduce three actual
physical scales: the heavy quark mass $m$, the soft scale of heavy quark
momentum inside the hadron $m v$ and ultrasoft scale of energy $m v^2$, which
are distinctly separated by a small parameter $v$ being the velocity of heavy
quark. After the matching with full QCD at a hard scale $\mu_{\rm hard}\sim m$,
in NRQCD the hard fields are integrated out, that results in the perturbative
Wilson coefficients of OPE in the effective theory, and we deal with the heavy
quarks interacting with the gluons at virtualities $\mu_{\rm fact,\, soft}$
about $m v$. In order to consider the heavy quark fields at lower $\mu$ up to
$m v^2$ we should introduce the effective lagrangian of pNRQCD, where the soft
fields are integrated out, and we deal with the potential interaction of heavy
quarks and the ultrasoft external gluon fields in the framework of multipole
expansion. The matching of pNRQCD with NRQCD takes place at a scale $\mu_{\rm
fact}\sim m v$. Recently, the effective theory of vNRQCD was formulated in
\cite{vNRQCD}, using the velocity renormalization group \cite{vRG} to match the
vNRQCD operators with the full QCD at a scale about $m$ with the single-step
evolution to a soft scale, which can be either $m v$ or $m v^2$. The current
status of vNRQCD provides us with the one-loop matching of heavy quark
potential to order $v^2$, i.e. up to spin-dependent $1/m^2$ terms, which are
beyond the current consideration. Therefore, we concentrate our discussion on
pNRQCD.

The pNRQCD lagrangian has the following form:
\begin{eqnarray}                         
& & {\cal L}_{\rm pNRQCD} = {\rm Tr} \,\Biggl\{ {\rm S}^\dagger \left(
i\partial_0 
- {{\bf P}^2\over 4m}- {{\bf p}^2\over m} +{{\bf p}^4\over 4m^3}
- V_s(r) - {V_s^{(1)} \over m}- {V_s^{(2)} \over m^2}+ \dots  \right) {\rm S}
\nonumber
\\
&&
\qquad 
+ {\rm O}^\dagger \left( iD_0 - {{\bf P}^2\over 4m} - {{\bf p}^2\over m}+{{\bf
p}^4\over 4m^3} 
- V_o(r) - {V_o^{(1)} \over m}- {V_o^{(2)} \over m^2}+\dots  \right) {\rm O}
\Biggr\}
\label{pnrqcdph}\\
& &\qquad + g V_A ( r) {\rm Tr} \left\{  {\rm O}^\dagger {\bf r} \cdot {\bf E}
\,{\rm S}
+ {\rm S}^\dagger {\bf r} \cdot {\bf E} \,{\rm O} \right\} 
+ g {V_B (r) \over 2} {\rm Tr} \left\{  {\rm O}^\dagger {\bf r} \cdot {\bf E}
\, {\rm O} 
+ {\rm O}^\dagger {\rm O} {\bf r} \cdot {\bf E}  \right\}  
\nonumber\\
& &\qquad- {1\over 4} F_{\mu \nu}^{a} F^{\mu \nu \, a}\,,\nonumber
\end{eqnarray}
where ${\bf P}$ is the momentum associated to the centre-of-mass coordinate. In
Eq. (\ref{pnrqcdph}) the  $1/m$ corrections to $V_A$, $V_B$ and to pure gluonic
operators as well as the higher order terms in the multipole expansion are not
displayed. To the leading order the singlet and octet operators $S$, $O$ are
represented by the appropriate products of nonrelativistic heavy quark and
antiquark spinors. The matching of $S$ and $O$ operators with the NRQCD spinors
was done in \cite{pNRQCD} up to three loops for both the potentials and the
normalization factors in OPE. In this lagrangian the singlet and octet
potentials $V_s(r)$ and $V_o(r)$ are treated as the corresponding Wilson
coefficients in front of bilinear forms in $S$ and $O$ to the leading order in
$1/m$. In ref. \cite{pNRQCD} the authors shown that this definition of static
quark potential is consistent with the definition in terms of Wilson loop
(\ref{def_WL}).

The other result of pNRQCD is the cancellation of renormalon
ambiguity in the sum of heavy quark pole masses and the potential up to two
loops, which is a confirmation of general consideration in QCD, that was first
derived in \cite{cancel}.

\setcounter{footnote}{0}

A new feature appears by the consideration of three-loop leading log matching
of $\rm V$ and $\overline{\rm MS}$ schemes. So, for the distance-dependent
running coupling the result reads off
\begin{eqnarray}
{\alpha}_{\Rsub V}(1/r^2, \mu) &=& \alpha_{\overline{\Rsub MS}}(1/r^2)
\left\{1+\left(a_1+ 2 {\gamma_E \beta_0}\right) {\alpha_{\overline{\Rsub
MS}}(1/r^2) \over 4\pi}\right.
\nonumber\\
&&+\left[\gamma_E\left(4 a_1\beta_0+ 2{\beta_1}\right)+\left( {\pi^2 \over 3}+4
\gamma_E^2\right) {\beta_0^2}+a_2\right] {\alpha_{\overline{\Rsub MS}}^2(1/r^2)
\over 16\,\pi^2}  \label{newpot}\\
&&\left. + {C_A^3 \over 12}{\alpha_{\overline{\Rsub
MS}}^3(1/r^2) \over \pi} \ln{ r\mu}\right\}, \nonumber
\end{eqnarray}
where the two-loop contribution was taken from \cite{Peter,Schroed} and it is
coincides with (\ref{Vr}), of course. However, the three-loop term leads to the
explicit dependence on the scale in the perturbative pNRQCD calculations, which
has to be expected from the general note on the infrared singularity observed
by Appelquist, Dine and Muzinich \cite{Su}, that was rederived in pNRQCD by
supplementing a certain infrared subtraction. This dependence was considered in
\cite{pNRQCD} for two cases, when the scales of confinement $\Lambda_{QCD}$ and
binding energy $m v^2$ have the arrangements {\it a}) $\Lambda_{QCD} \gg m
v^2$ or {\it b})  $m v^2 \gg \Lambda_{QCD}$. If {\it a}), the singlet
potential of static quarks suffers from the nonperturbative effects, and it can
be treated only after introduction of some model dependent terms coming from
the ultrasoft gluons, which form the gluon sea in the heavy quarkonium, so that
the sea has its excitations, and the characteristic excitation energy of
gluelumps should replace the scale $\mu$, that results in the scale-independent
nonperturbative potential\footnote{Possible non-potential terms are discussed
in \cite{pNRQCD}.}. If {\it b}), the potential is purely perturbative.
However, calculating the physical quantities such as the masses of bound
states, we have to take into account the contributions coming from the
perturbative ultrasoft gluons with the virtualities less than $\mu$, which can
produce a $\mu$-dependent shift of energy, that should be cancelled with the
$\mu$-dependence in the potential (\ref{newpot}) and, probably, in the heavy
quark masses. In both cases, the perturbative calculations of singlet
potential\footnote{We do not concern for the octet potential of static quarks
in the present consideration, though some qualitative conclusions could be
straightforwardly generalized from the siglet state to the octet one.}
explicitly indicate the necessity to take into account the gluon degrees of
freedom inside the heavy quarkonium. As was noted in \cite{pNRQCD}, apparently,
this feature is characteristic for the nonabelian theory (see the factor of
$C_A$ in front of log term in (\ref{newpot})).

To our opinion, this dependence of potential on the ultrasoft gluon fields (the
infrared singularity in terms of Appelquist, Dine and Muzinich) inside the
heavy quarkonium naturally indicates the formation of gluon string between the
heavy quarks at long distances. Indeed, expression (45) was derived under the
following arrangement of scales: $r\sim m v$, $m v^2 < \mu < m v$. So,
if we put
$$
\mu = \frac{u}{r} - \sigma\, r,
$$
with
$$
v < u < 1,\;\;\;{\rm and}\;\;\; \sigma \ll \frac{u}{m^2 v^2},
$$
then, perturbatively expanding in the small parameter $\sigma\, r$, we get the
linear correction to the potential in pNRQCD, so that
$$
\Delta V_{\rm pNRQCD} = \Delta k\cdot r,
$$
where 
$$
\Delta k = \frac{C_F C_A^3}{12\pi}\, \alpha_{\overline{\Rsub
MS}}^4\,\frac{\sigma}{u} \approx \frac{C_A^3}{12\pi}\, \alpha_{\overline{\Rsub
MS}}^3\,{\sigma},
$$
so that we have dropped the scale dependence of strong coupling constant, since
it is beyond the accuracy under study, and we have substituted the coulomb
relation for the quark velocity inside the bound state $u\approx C_F
\alpha_{\overline{\Rsub MS}}$ to the given order. Numerically, for the charmed
quarks this perturbative contribution could be of the order of $\Delta k \sim
0.1$ GeV$^2$. Thus, we can motivate the relation between the nonperturbative
string and the three-loop scale dependent term in the pNRQCD potential.

Indifferently of the arrangement
for the confinement and binding energy scales, the introduction of such the
string should remove the explicit dependence of full potential on the scale.
This has been done above by introduction of unified $\beta$ function of
coupling in the $\rm V$ scheme. This solution of the problem qualitatively
agrees with the consideration in pNRQCD, since, first, in the perturbative
regime the contribution of log term is negligibly small as we see for the
linear confining term of potential at short distances, and, second, at long
distances the nonperturbative confining term is essential, where the string
tension is the natural physical scale. In the static potential of QCD given
above we do not consider possible `nontrivial' excitations with the broken
string geometry, where the break point moves on the string with the speed of
light. Such the excitations would correspond to hybrid states with the
gluelumps. Thus, we find that the QCD potential of static quarks in the form
offered in the present paper has no conflicts with the current status of
pNRQCD.

However, to our opinion the problem can be more deep. The static potential,
introduced by the Wilson loop, is renormalization group invariant, and it does
not contain any separation between the potential gluons and the ultrasoft
gluons forming the sea, since it gives the total energy of dynamical fields. In
contrast, the pNRQCD introduced the singlet potential as the Wilson coefficient
in front of four quark operator, so that it intrinsically operates with the
separation of potential and sea, as well as the nonrelativistic quarks, which
act as sources, so that some gluons with virtualities greater than $\mu$ are
considered as emitted, while others with virtualities less than $\mu$ are
included into the origin of sources, and the gluons with virtualities about $m
v$ mediate the potential interaction. Generally, this separation of heavy
quarks, potential gluons and sea gluons in the operator product expansion can
involve nonzero anomalous dimensions for the singlet pNRQCD-potential, say.
This fact does not contradict with the OPE basis, but it reflects the point
that the static potential of Wilson loop generally differs from the
pNRQCD-potential. In addition, the ultrasoft gluon sea introduced in pNRQCD in
terms of multipole interaction with local external chromoelectric and
chromomagnetic fields is not a local object, indeed.

A note should be done on the linear confining term of potential. In
\cite{pNRQCD} a model of infrared behaviour was used, so that at long distances
between the heavy quarks the ultrasoft correction was derived in the form of
constant energy shift $\delta V_0$ and quadratic term $\sigma_2 r^2$. The
corresponding conclusion was drawn to stress that the linear term could appear
in a more complicated case of infrared behaviour. We show in the previous
sections how this confinement regime can be reached.

Recently, several papers \cite{ptBC,ptQQ} were devoted to the calculations of
ground states in the heavy quarkonia in the way, combining the pNRQCD potential
with the nonperturbative corrections to the binding energy as they produced by
the multipole expansion of QCD \cite{VolLeut} in the form of pNRQCD explicitly
shown in (\ref{pnrqcdph}). Ref. \cite{ptBC} does not strictly estimate the
gluon condensate effects in the multipole expansion, and it presents purely
perturbative results. It follows the perturbative ground state mass technique
as a mass definition that leads to the cancellation of the $u=1/2$ renormalon
that was considered in the approach of upsilon expansion introduced by Hoang et
al. in \cite{H2}. So, in \cite{ptBC} the perturbative mass of $B_c$ meson was
calculated on the base of perturbative expansion for the static
potential with the leading approximation in the form of coulomb wave functions.
As we see above the perturbative potential suffers from the renormalon
ambiguity. In order to remove this dependence on the choice of scale $\mu$ in
the potential, the authors of \cite{ptBC} calculated the masses of $J/\Psi$ and
$\Upsilon$ in the same technique at the same point $\mu$ and inverted the
problem on the heavy quark masses by equalizing the perturbative masses of
ground states in the charmonium and bottomonium to the measured values. This
procedure leads to the $\mu$-dependent pole masses of heavy quarks as expressed
by the series in $\alpha_s(\mu)$. We expect that such the procedure could
cancel the renormalon with the accuracy about 50 MeV in the mass of hadron. As
a results, the perturbative mass of $B_c$ has quite a stable value 
\begin{equation}
M_{\rm pert}(B_c) = 6326^{+29}_{-9}\;{\rm MeV},
\label{pertBc}
\end{equation}
in the range of $1.2 < \mu < 2.0$ GeV, which should be compared with the
results in Table \ref{bc} and the range of $\mu$ described above in the study
of matching the perturbative potential with the full QCD one. The authors of
\cite{ptBC} did not present the $\mu$-dependent heavy quark masses.
Nevertheless, due to the almost coinciding estimates of $B_c$ mass in
(\ref{pertBc}) and Table \ref{bc}, we expect that this dependence should be
given by the form of $\delta V(\mu)$.

In refs. \cite{ptQQ} the same technique for the perturbative contribution with
the account for both the gluon condensate corrections in the multipole
expansion of QCD and a small $\alpha_s^5 \log \alpha_s$ term, was used to
extract the heavy quark masses. The authors determined the `pole' mass, which
is scale dependent, indeed, by putting $\mu = C_F \alpha_s m_Q$ in the
potential. As we understand, they introduced the mass suffered from the
renormalon and got
$$
m_b = 5022\pm 58\; {\rm MeV},
$$
which is greater than we determine in the current presentation. The reason is
quite evident. It is the energy shift $\delta V(\mu)$. The running
$\overline{\rm MS}$ mass quoted in \cite{ptQQ} is about 260 MeV greater than we
find in the same order in $\alpha_s$ for the relation between the pole and
running masses. The difference becomes unessential by using the three-loop
matching of the masses in \cite{ptQQ}, however, the same correction will also
decrease the value obtained in the spectroscopy with the full QCD potential.
Thus, to our opinion the values of heavy quark masses given in \cite{ptQQ}
should be kept with a large care.

Finally, in \cite{melles} the dependence of potential on the finite heavy quark
masses was considered. This dependence is due to the smooth variation of number
of active flavors in the expressions for the coefficients of perturbative
$\beta$ function as well as in the matching coefficients of $\alpha_{\Rsub
V}$. As we have described above we use the step-like change of active flavor
number, which infers implicit model-dependence, which is practically
unavoidable in the case under study.

As for the lattice simulations in QCD for the relevant problem, the review can
be found in ref. \cite{bali}. We emphasize only that the lattice potential of
static quarks is close to what is given by the Cornell model. A modern review
of phenomenological potential models can be found in lectures \cite{nora}. The
finite mass effects in the nonrelativistic bound states was recently
considered at nest-to-leading order in \cite{manetal} and \cite{add}. A
next-to-next-to-leading order analysis of light quark mass effects in the heavy
nonrelativistic $Q\bar Q$ systems was given in \cite{H3}. Some applications of
pNRQCD to the heavy quarkonia were done in \cite{usoft}.

\subsection{Leptonic constants}

In the static approximation for the heavy quarks the calculation of leptonic
constants for the heavy quarkonia with the two-loop accuracy involves the
matching of leptonic currents in NRQCD with the currents of full QCD,
$$
J_\nu^{QCD}= \bar Q \gamma_\nu Q, \;\;\; {\cal J}_\nu^{NRQCD} = \chi^\dagger
\sigma_\nu^\perp \phi,
$$
with the relativistic quark fields $Q$ and their nonrelativistic two-component
limits of antiquark $\chi$ and quark $\phi$, $\sigma_\nu^\perp= \sigma_\nu
-v_\nu (\sigma \cdot v)$, and $v$ is the four-velocity of heavy quarkonium, so
that
$$
J_\nu^{QCD} = {\cal K}(\mu_{\rm hard}; \mu_{\rm fact})\cdot {\cal
J}_\nu^{NRQCD},
$$
where the scale $\mu_{\rm hard}$ determines the normalization point for the
matching of NRQCD with full QCD, while $\mu_{\rm fact}$ refers to the point of
perturbative calculations in NRQCD. Using the matching of potential for the
static quarks in QCD with the two-loop perturbative potential, we argue that
the most appropriate choice of scale relevant to the charmonium and bottomonium
is
\begin{equation}
\mu_{\rm fact} = \mu_{\rm soft} = 1.3 - 2\;{\rm GeV.}
\label{soft}
\end{equation}
For the heavy quarkonium composed by quarks of the same flavour the Wilson
coefficient ${\cal K}$ is known up to the two-loop accuracy
\cite{HT,mel,bensmir,melch}
\begin{equation}
{\cal K}(\mu_{\rm hard}; \mu_{\rm fact}) = 1 -\frac{8}{3}
\frac{\alpha_s^{\overline{\Rsub MS}}(\mu_{\rm hard})}{\pi}+
\left(\frac{\alpha_s^{\overline{\Rsub MS}}(\mu_{\rm
hard})}{\pi}\right)^2 c_2(\mu_{\rm hard}; \mu_{\rm fact}),
\label{kfact}
\end{equation}
and $c_2$ is explicitly given in \cite{bensmir,melch}. The additional problem
is the convergency of (\ref{kfact}) at the fixed choice of scales. So, putting
$\mu_{\rm hard} = (1-2) m_{b}$ and (\ref{soft}) we find a good convergency of
QCD corrections for the bottomonium and estimate its leptonic constant defined
by
$$
\langle 0| J_\nu^{QCD} |\Upsilon,\lambda \rangle = \epsilon_\nu^\lambda
f_{\Upsilon} M_{\Upsilon},
$$
where $\lambda$ denotes the polarization of vector state $\epsilon_\nu$, so
that
$$
f_{\Upsilon} = 685\pm 30\; {\rm MeV,}
$$
while the experimental value is equal to $f^{\rm exp}_{\Upsilon} = 690\pm 13\;
{\rm MeV}$ \cite{PDG}.

\begin{figure}[ph]
\setlength{\unitlength}{1mm}
\begin{center}
\begin{picture}(100,90)
\put(5,5){\epsfxsize=9cm \epsfbox{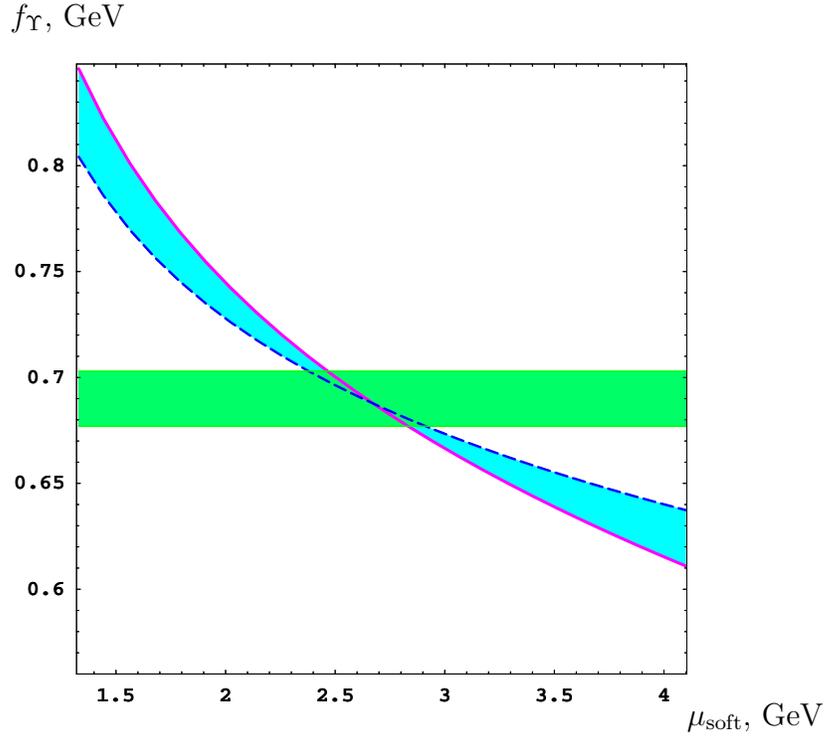}}
\put(95,5){$\mu_{\rm soft}$, GeV}
\put(5,98){$f_{\Upsilon}$, GeV}
\end{picture}
\end{center}

\caption{The value of leptonic constant for the vector ground state of
bottomonium versus the soft scale. The dashed line represents the choice of
$\mu_{\rm hard} = 2 m_b$, while the solid line does $\mu_{\rm
hard} = m_b$. The horizontal shaded band gives the
experimental limits.} 
\label{fups}
\end{figure}

As we can see in Fig. \ref{fups} the variation of hard scale in broad limits
leads to existence of stable point, where the result is slowly sensitive to
such the variation. The stability occurs at $\mu_{\rm soft}\approx 2.6$ GeV,
where the perturbative potential is still close to the potential of static
quarks at the distances characteristic for the $1S$-level of $\bar b b$.

The estimate of leptonic constant for the charmonium $J/\psi$ is more sensitive
to the choice of factorization scale. Indeed, the size of this system, $\langle
r_{\bar c c(1S)}\rangle \approx 0.42$ fm, makes more strict constraints on
$\mu_{\rm fact} \approx 1.3-1.5$ GeV, since at higher scales the perturbative
potential significantly deviates from the potential of static quarks in QCD in
the region of bound $\bar c c$ states, while at lower scales the perturbative
potential in two loops does not match the QCD potential in all of the form.

\begin{figure}[ph]
\setlength{\unitlength}{1mm}
\begin{center}
\begin{picture}(100,97)
\put(5,5){\epsfxsize=9cm \epsfbox{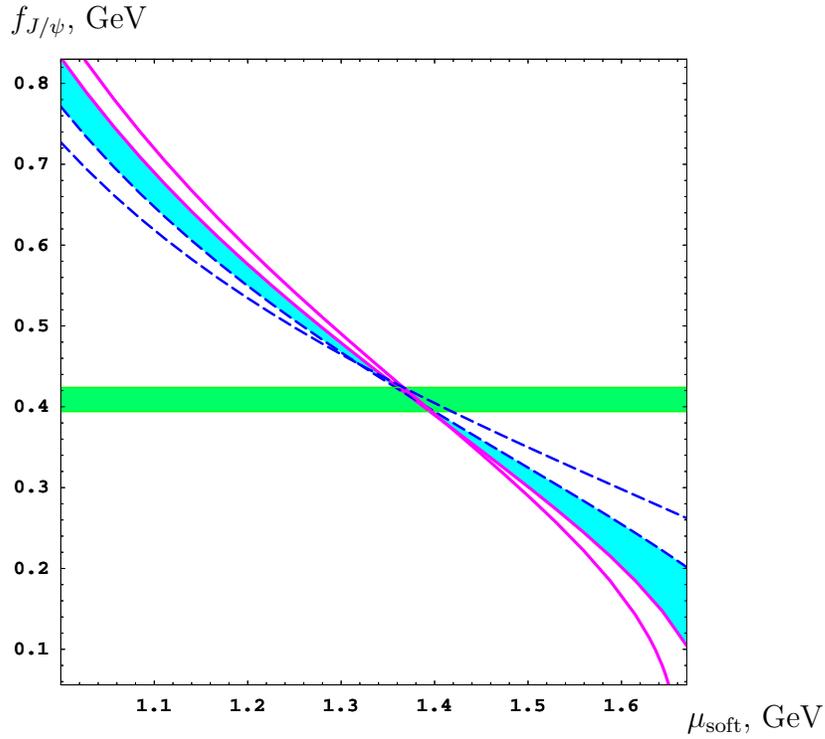}}
\put(95,5){$\mu_{\rm soft}$, GeV}
\put(5,98){$f_{J/\psi}$, GeV}
\end{picture}
\end{center}

\caption{The value of leptonic constant for the vector ground state of
charmonium versus the soft scale. Shaded region is restricted by the dashed
line representing the choice of $\mu_{\rm hard} = 1.07 m_c$, and the solid line
with $\mu_{\rm hard} = 0.93 m_c$. The horizontal shaded band gives the
experimental limits. The additional curves give $\mu_{\rm hard} = 1.26 m_c$
(dashed line) and $\mu_{\rm hard} = 0.87 m_c$ (solid line).} 
\label{fpsi}
\end{figure}
Another problem is the energy shift $\delta V(\mu)=1.0-1.2$ GeV, which
essentially renormalizes the pole mass of charmed quark: $m_c^{\rm pole}
=1.968-2.068$ GeV. This shift does not perturb the mass of the ground state,
but it is significant for the value of wave function at the origin. So,
following the well-adjusted scaling relation for the leptonic constants
\cite{scale}, we put $P(\mu) =\kappa \Psi(0)\; m_c^{\rm pole}(\mu)/m_c$ and use
it in the calculations of the leptonic constant\footnote{Solving the
Schr\"odinger equation with the shifted masses and potential, we check that
this mass dependence of wave function is valid with the accuracy better than
6\%, so we put $\kappa=0.95$.}. We get
$$
f_{J/\psi} = 400\pm 35\; {\rm MeV,}
$$
to compare with the experimental value $f^{\rm exp}_{J/\psi} = 409\pm 15\; {\rm
MeV}$.

In Fig. \ref{fpsi} we see that again the stability point can be reached in the
variation of $\mu_{\rm hadr}$ at reasonable value of $\mu_{\rm soft}\approx
1.35$ GeV. However, the stability takes place in the narrow region of $\mu_{\rm
hadr}$ close to the charm quark mass.

At present, the matching condition for the heavy quarkonium composed by the
quarks of different flavors, $\bar b c$, is known to one loop, only
\cite{scale,bf}. So, for the pseudoscalar state we have
\begin{equation}
{\cal K}(\mu_{\rm hard}; \mu_{\rm fact}) = 1 -
\frac{\alpha_s^{\overline{\Rsub MS}}(\mu_{\rm hard})}{\pi}
\left(2-\frac{m_b-m_c}{m_b+m_c}\ln \frac{m_b}{m_c}\right),
\label{kfactbc}
\end{equation}
which is independent of the factorization scale. The matching of perturbative
potential to the one-loop accuracy with the QCD potential of static quarks at
$r\sim 0.3 - 0.4$ fm relevant to the ground state of $B_c$ meson \cite{rev}, is
rather questionable, since the deviation in the forms of potentials is quite
sizable. In addition we have to pose $\mu_{\rm fact} = \mu_{\rm hard}$, because
we cannot distinguish these scales, while the nonzero anomalous dimension to
two loops is not taken into account. Nevertheless, we can put $\mu_{\rm
hard}=1.3-1.8$ GeV and neglect $\delta V$, which is beyond the actual control
in the one-loop accuracy. Indeed, as we see in Fig. \ref{shift} the one loop
value of energy shift for the matching of perturbative and QCD potentials is
quite small at the large virtualities about 2 GeV, and it can be neglected,
while at smaller virtualities the form of perturbative potential is close to
that of given by QCD only in the short range of distances $r = 0.1-0.25$ fm,
hence, the results on the matching are not reliable for the extracting the
heavy quark masses from the parameters of bound states. So, we estimate
$$
f_{B_c} = 400\pm 45\; {\rm MeV,}
$$
to compare with the estimates in the SR, where $f^{\rm SR}_{B_c} = 400\pm 25\;
{\rm MeV}$ \cite{scale,fbc}.
\begin{table}[th]
\caption{The ratios of leptonic constants for the heavy quarkonia as predicted
in the present paper (K$^2$O) in comparison with the experimental data.}
\begin{center}
\begin{tabular}{|l|r|r||l|r|r|}
${f^2_{\psi(nS)}}/{f^2_{\psi}}$ & QCD, K$^2$O & exp. &
${f^2_{\Upsilon(nS)}}/{f^2_{\Upsilon}}$ & QCD, K$^2$O & exp. \\
\hline
2S & 0.55 & $0.48\pm 0.07$ & 2S & 0.47  & $0.47\pm 0.03$ \\
3S & 0.32 & $0.25\pm 0.06$ & 3S & 0.34  & $0.36\pm 0.02$ \\
\end{tabular}
\end{center}
\label{excit}
\end{table}

Finally, we present the ratios of leptonic constants for the excited
$nS$-levels of $\bar b b$ and $\bar c c$ in Table \ref{excit} in comparison
with the experimental data. We see that the predictions are in a good agreement
with the measured values. For completeness, we predict also the constant of
$2S$-level in the $\bar b c$ system
$$
f_{B_c(2S)} = 280\pm 50\; {\rm MeV,}
$$
which agrees with the scaling relation \cite{scale}.

Thus, we have analyzed the estimates following from the potential of static
quarks in QCD for the masses of quarks and heavy quarkonia as well as for the
leptonic constants, and found both the good agreement with the experimental
data available and the consistency with the QCD sum rules.

\section{Conclusion}

We have derived the potential of static heavy quarks in QCD on the base of
known limits at short and long distances: the asymptotic freedom to the three
loop accuracy and the confinement regime. The inputs of potential are the
coefficients of perturbative $\beta$ function, the matching of $\overline{\rm
MS}$ scheme with the V scheme of potential, the normalization of running
coupling constant of QCD at $\mu^2 = m_Z^2$ and the slope of Regge
trajectories, determining the linear term in the potential. Thus, the approach
by Buchm\"uller and Tye has been modified in accordance with the current status
of perturbative calculations.

In the static limit the two-loop improvement of coulomb potential results in
the significant correction to the $\beta$ function for the effective charge,
$\Delta \beta/\beta \sim 10\%$ as shown in Fig. \ref{dif}. This correction is
important for the determination of critical values of charge, i.e. the value in
the intermediate region between the perturbative and nonperturbative regimes.
Moreover, the two-loop matching condition and the three-loop running of
coupling constant normalized by the data at the high energy of $m_Z$ determine
the region of energetic scale for changing the regimes mentioned above. This
scale strongly correlates with the data on the mass spectra of heavy quarkonia.
So, it is connected with the splitting of masses between the $1S$ and $2S$
levels. We stress that the consistent consideration of two-loop improvement
gives the appropriate value of effective coulomb coupling constant as it was
fitted in the Cornell model of potential. This is achieved in the present paper
in contrast to the one-loop consideration by Buchm\"uller and Tye, who found
the value of $\Lambda_{QCD}$ inconsistent with the current normalization at
high energies. So, the two-loop improvement gives the correct normalization of
effective coulomb exchange at the distances characteristic for the average
separation between the heavy quarks inside the heavy quarkonium and determines
the deviations at short distances $r < 0.08$ fm (see Fig. \ref{diffCornell}),
that is important in the calculations of leptonic constants related with the
wave functions at the origin.

Other corrections to the potential of heavy quarks are connected with the
finite mass effects and cannot be treated in the framework of static
approximation. For example, the spin-dependent forces, relativistic corrections
and specific non-abelian potential terms\footnote{They have the form of
$\alpha_s^2/r^2$ with the factor given by the inverse heavy quark masses.} in
the heavy quarkonium should be taken in the analysis of mass spectra. A
magnitude of leading nonstatic corrections can be evaluated by the
characteristic shifts of levels due to the hyper-fine splitting of $S$-wave
levels in the heavy quarkonia\footnote{The splitting is about 100 MeV or
less.}.  So, we conservatively evaluate the uncertainty of heavy quark mass
analysis $\delta m \simeq 80$ MeV.

Thus, the non-abelian term of potential $\alpha_s^2/r^2$, say, has the factors
in the form of $1/m_Q$, and it is equal to zero in the static limit $m_Q\to
\infty$, while the uncertainty in the heavy quark masses due to the omission of
such the terms is estimated in the paragraph above. Formally, if we consider
the perturbation theory for the calculation of bound state levels in the heavy
quarkonium with the coulomb functions taken as the leading approximation, which
is not a scope of our consideration, then the mentioned non-abelian potential
contributes in the same order in $\alpha_s$ as the two-loop corrections to the
matching of perturbative static potential $\sim \alpha_s^4$, since the
averaging of $1/r^2$ results in $\alpha_s^2 m_Q^2$ factor. However, the
two-loop effects are important for the consistent consideration of static
potential and the high energy normalization, i.e. these corrections are
significant in the running of effective charge in the potential from the high
energies to the scale relevant to the heavy quark bound states even in the
static limit, while the nonstatic contributions can be consistently neglected
in the numerical analysis. We see that our consideration is consistent in the
static approximation, which we have addressed in the present paper.

The matching of two loop perturbative potential with the QCD potential of
static quarks has been performed to get estimates of heavy quark masses, which
can be compared with the results of QCD sum rules. A good agreement between two
approaches has been found.

The recent determinations of heavy quark masses in Refs. \cite{ben,mel,hoang}
were done in the framework of QCD sum rules, which is a systematic approach,
indeed. It is based on the separation of short-distance region from the
nonperturbative effects at some values of parameters defining the scheme of
calculations in the sum rules. In this approach the nonperturbative terms are
given in the form of quark-gluon condensates contributing with corresponding
short-distance Wilson coefficients, so that as was shown in \cite{mbv}, a
numerical contribution of gluon condensate term in the sum rules is negligibly
small in comparison with the perturbative part. However, it would be incorrect
to think that these explicit contributions suppressed in some region of
parameters are the only terms caused by the nonperturbative infrared dynamics
of QCD. Indeed, neglecting the condensate terms, we find that the perturbative
correlators suffer from the renormalon ambiguity, which implies that the
perturbative expansion in series of $\alpha_s$ is asymptotic, and the summation
of series depends on a method used. The physical reason for such the divergency
and ambiguity is the infrared singularity in the QCD coupling constant. This
singularity is regularized by introducing the threshold mass parameters free of
renormalon. Such the approach is independent of any assumptions on the gluon
condensate, since, generally the pole mass renormalon and the gluon condensates
are different issues.

So, the perturbative pole mass used in the QCD sum rules is not
well defined quantity, and some relevant quantities are introduced in Refs.
\cite{ben,mel,hoang}. These quantities are constructed from the perturbative
pole mass of heavy quark with specific infrared subtractions, which are treated
independently of the quark-gluon condensates. These constructions are
author-dependent. Though the authors of subtracted masses gave some physical
motivations, which are more or less strict, but justified. These infrared
subtractions imply the introduction of infrared regulators. 

In the present paper the unified $\beta$ function for the effective charge in
the potential is considered, and its definition supposes the infrared
stability. Thus, we see that the analysis of heavy quark masses in both the QCD
sum rules and potential approach involves the consideration of relevant effects
caused by the infrared dynamics of QCD, though the explicit constructive
procedures are certainly different, but they have similar inherent
uncertainties.

The calculated mass spectra of heavy quarkonia and the leptonic constants of
vector $nS$-levels are in agreement with the measured values. The
characteristics of $B_c$ meson have been predicted.

The authors are grateful to prof. A.K.Likhoded and A.L.Kataev for stimulating
discussions and A.A.Pivovarov for clarifying the results of QCD sum rules on
the $b$-quark mass. We thank Dr. Antonio Vairo for valuable remarks, references
and explanations concerning the approach of pNRQCD, and for discussions.

This work is in part supported by the Russian Foundation for Basic Research,
grants 01-02-99315, 01-02-16585 and 00-15-96645. The work of A.I.Onishchenko
was supported, in part, by International Center of Fundamental Physics in
Moscow, International Science Foundation, and INTAS-RFBR-95I1300 grants.


\end{document}